\documentclass[aps,pra,10pt,twocolumn,nofootinbib]{revtex4-2}
\usepackage[cm]{fullpage}
\usepackage{microtype}
\usepackage[T1]{fontenc}
\usepackage{times}
\usepackage{amsfonts,amsmath,amssymb,mathrsfs}
\usepackage{graphicx}
\usepackage{xcolor}
\usepackage[pdfstartview=FitH,colorlinks=true,citecolor=blue,linkcolor=blue,urlcolor=blue]{hyperref}

\begin{document}

\title{Critical velocity for superfluidity in the one-dimensional mean-field regime: \texorpdfstring{\\}{Lg} From matter to light quantum fluids}
\author{J. Huynh}
\author{M. Albert}
\email{mathias.albert@inphyni.cnrs.fr}
\author{P.-\'E. Larr\'e}
\email{pierre-elie.larre@inphyni.cnrs.fr}
\affiliation{Universit\'e C\^ote d'Azur, CNRS, Institut de Physique de Nice (INPHYNI), France}

\begin{abstract}
We determine in a nonperturbative way the critical velocity for superfluidity of a generic quantum fluid flowing past a localized obstacle in the one-dimensional mean-field regime. We get exact expressions in the narrow- and wide-obstacle limits and interpolate them numerically using an original relaxation algorithm for the stationary problem. The existence of a Josephson-type critical current across a very high and slowly varying obstacle is discussed. Particle losses, if present, are treated within an adiabatic approach of the dynamics giving results in excellent agreement with full numerics. Relevant for experiments with quantum fluids of matter, of mixed matter-light, and of light, our study paves the way for further nonperturbative investigations in higher dimensions and beyond mean-field theory.
\end{abstract}

\maketitle

\section{Introduction}
\label{Sec:Introduction}

Superfluidity, one of the most striking manifestations of the collective behavior of matter at low temperature, is the ability of a quantum fluid to flow along a narrow capillary, beside the wall of a bucket, or past a localized obstacle without any loss of kinetic energy \cite{Leggett1999, Balibar2007, Pitaevskii2016}. This spectacular phenomenon consistently occurs below a specific flow speed, the so-called critical velocity for superfluidity. It was evidenced for the first time by Kapitza \cite{Kapitza1938}, Allen, and Misener \cite{Allen1938} in experiments with liquid helium-4 below the $\lambda$-point, later followed by no less pioneering experiments with liquid helium-3 in its paired phase \cite{Osheroff1972a, Osheroff1972b}, ultracold atomic Bose \cite{Raman1999, Onofrio2000, Raman2001, Engels2007, Neely2010, Dries2010, Desbuquois2012} and Fermi \cite{Miller2007} gases, exciton-polariton condensates in semiconductor optical microcavities \cite{Amo2009, Lerario2017}, and more recently laser beams which behave as quantum fluids when propagating in cavityless nonlinear dielectrics \cite{Vocke2016, Michel2018, Eloy2021}.

In the wake of the discovery of superfluidity in a flow of helium-4 along a capillary \cite{Kapitza1938, Allen1938}, Landau theorized the phenomenon \cite{Landau1941a, Landau1941b}, notably by being interested in the condition under which the nucleation of an elementary excitation by friction of the fluid with the capillary walls becomes energetically favorable. This condition concerns the flow velocity $\boldsymbol{v}$, which Landau showed that the norm has to be larger than the ratio of the energy $\varepsilon(\boldsymbol{p})$ to the norm of the momentum $\boldsymbol{p}$ of an elementary excitation in the fluid at rest. A frictionless flow along the capillary is thus possible as long as $|\boldsymbol{v}|<\min_{\boldsymbol{p}}\varepsilon(\boldsymbol{p})/|\boldsymbol{p}|$, the right-hand side defining the critical velocity for superfluidity. This is the celebrated Landau criterion for superfluidity \cite{Pitaevskii2016}, which is sufficiently general to be applied to quantum fluids other than superfluid helium-4 and to flow configurations other than along a channel, for instance when the fluid hits a localized obstacle. Despite its elegant simplicity---indeed, it only requires to know the dispersion relation of the elementary excitations of the quantum fluid at rest---, Landau's argument often overestimates the critical velocity for frictionless motion, as noticed by Landau himself in the context of helium-4 superfluidity \cite{Landau1947}. This results from the fact that it is by essence perturbative. Indeed, in a quantum fluid like superfluid helium-4, energy losses manifest not only through the emission of elementary excitations but also through the generation of nonlinear excitations such as quantized vortices \cite{Pitaevskii2016}, which are stimulated at flow speeds smaller than the Landau critical velocity for superfluidity. This was first understood by Feynman \cite{Feynman1955}, who derived a speed threshold for vortex proliferation much closer to the experimental critical velocity for helium-4 superfluidity than Landau's theoretical counterpart.

Since Landau's and Feynman's seminal works, the precise determination of the velocity marking the onset of superfluidity in a flowing quantum fluid continues to challenge theoreticians. Boosted by related experiments with atomic Bose-Einstein condensates \cite{Raman1999, Onofrio2000, Raman2001, Engels2007, Neely2010, Dries2010, Desbuquois2012}, numerous theoretical investigations of the critical velocity for superfluidity in a nonlinear Schr\"odinger \cite{Pitaevskii2016} flow past a localized obstacle emerged. Many of them are carried out in two \cite{Frisch1992, Josserand1997, Josserand1999, Huepe2000, Stiessberger2000, Rica2001, Sasaki2010, Pinsker2014, Singh2017} and three \cite{Berloff2000, Stiessberger2000, Weimer2015, Singh2016} dimensions (2 and 3D) (see also Ref.~\cite{Pigeon2021} for an extension of the results of Ref.~\cite{Frisch1992} to resonantly driven exciton-polariton condensates). Due to the nonintegrability of the equations modeling the hydrodynamics of the superfluid in these dimensions, these studies mostly rely on numerical simulations and report analytical results only in a very few limiting cases, often when the obstacle consists in an impenetrable disk or ball whose radius is very large compared to the healing length \cite{Pitaevskii2016} of the unperturbed quantum fluid. In one dimension (1D) instead, exact solutions exist to the tunneling of a nonlinear Schr\"odinger fluid through a potential barrier and analytical expressions for the corresponding critical velocity for superfluidity can be obtained in various obstacle geometries \cite{Sols1994, Zapata1996, Hakim1997, TarasSemchuk1999, Leboeuf2001, Pavloff2002, Abid2003, Danshita2006, Danshita2007, Albert2008, Albert2009, Watanabe2009, Albert2010, Arabahmadi2021a, Arabahmadi2021b} (see also Refs.~\cite{Larre2012, Larre2015a} for related works on light superfluidity in incoherently pumped exciton-polariton condensates and in paraxial nonlinear optics). In this dimension, the breakdown of superfluidity typically manifests through the repeated emission of solitons \cite{Pitaevskii2016}, which are the 1D equivalents of the vortices discussed above.

In this paper, we determine in a nonperturbative way the critical velocity for superfluidity of a generic quantum fluid flowing past a localized obstacle in the 1D mean-field regime. Our model of superflow, detailed in Sec.~\ref{Sec:SuperfluidHydrodynamics}, relies on a generalization of the 1D nonlinear Schr\"odinger equation to any local self-interaction potential increasing with the fluid density. This makes it possible to describe various superfluid systems ranging from ultracold atomic Bose and Fermi gases \cite{Pitaevskii2016} to exciton-polariton condensates in semiconductor optical microcavities \cite{Carusotto2013} and fluids of light \cite{Leboeuf2010, Larre2015b, Vocke2016, Santic2018, Michel2018, Cherroret2018, Glorieux2018, Eloy2021}, as reviewed in Appendix~\ref{App:A}. The localized obstacle we consider is described by a smooth, even, and repulsive or attractive potential characterized by a single extremum and a range of arbitrary values, i.e., not restricted to the limits imposed by the perturbative Landau criterion. Building atop Refs.~\cite{Hakim1997, Leboeuf2001, Pavloff2002}, we derive analytical results for the critical velocity, first in the narrow-obstacle limit in Sec.~\ref{Sec:NarrowObstacle} and Appendix~\ref{App:B}, and then in the inverse, wide-obstacle limit in Sec.~\ref{Sec:WideObstacle} and Appendix~\ref{App:C}. Following Ref.~\cite{Dalfovo1996}, the occurence of a Josephson-type critical current across a very high and slowly varying obstacle is discussed at the end of Sec.~\ref{Sec:WideObstacle} with Appendix~\ref{App:D} as support. The obstacle of arbitrary width is numerically treated in Sec.~\ref{Sec:ObstacleOfArbitraryWidth} using an original relaxation algorithm for the stationary problem. Especially important in experiments on light superfluidity, we account for particle losses in the model. We treat them in Sec.~\ref{Sec:EffectOfParticleLosses} within an adiabatic approach \cite{Larre2017, Fontaine2020, Eloy2021} of the time evolution of the wavefunction. This provides results for the critical velocity in very good agreement with full numerics. We finally conclude and give perspectives to the present work in Sec.~\ref{Sec:ConclusionAndPerspectives}.

\section{Superfluid hydrodynamics}
\label{Sec:SuperfluidHydrodynamics}

A 1D quantum fluid of particles of mass $m$ flows in the positive-$x$ direction. In the mean-field regime, information on its density $n(x,t)$ ($t$ is time) and velocity $v(x,t)=\hbar\theta_{x}(x,t)/m$ ($\hbar$ is the reduced Planck constant) is encapsulated in a complex wavefunction $\psi(x,t)=n^{1/2}(x,t)\exp[i\theta(x,t)]$ whose dynamics is supposed to be ruled by the following generalized nonlinear Schr\"odinger equation:
\begin{equation}
\label{Eq:GNLSE}
i\hbar\psi_{t}=-\frac{\hbar^{2}}{2m}\psi_{xx}+U(x)\psi+g(|\psi|^{2})\psi-\frac{i\hbar\gamma}{2}\psi.
\end{equation}
The flow is here constrained by an obstacle described in Eq.~\eqref{Eq:GNLSE} by a smooth, even, and repulsive (attractive) potential $U(x)=U_{0}f(|x|/\sigma)$ which attains its single positive maximum (negative minimum) $U_{0}$ at $x=0$ and which is localized, i.e., which vanishes as $|x|\gg\sigma$, with $\sigma$ being its typical range. For concreteness, the Gaussian potential $U(x)=U_{0}\exp(-x^{2}/\sigma^{2})$ is of this type. The fluid is also subjected to a self-interaction described in Eq.~\eqref{Eq:GNLSE} by the local nonlinear term $g(|\psi|^{2}=n)\psi$, where the potential $g(n)$ is an increasing function of the density $n$. Finally, the last term in the right-hand side of Eq.~\eqref{Eq:GNLSE} describes linear particle losses with constant rate $\gamma$.

Equation~\eqref{Eq:GNLSE} governs the dynamics of a wide variety of systems including, e.g., zero-temperature atomic Bose-Einstein condensates and Fermi superfluids \cite{Pitaevskii2016} in highly anisotropic traps, exciton-polariton condensates \cite{Carusotto2013} in wire-shaped semiconductor optical microcavities, and fluids of light \cite{Leboeuf2010, Larre2015b, Vocke2016, Santic2018, Michel2018, Cherroret2018, Glorieux2018, Eloy2021} in 1D geometries. We refer the reader to Appendix~\ref{App:A} for further details. Although totally generic, the results we establish in the following will be often examplified in the case of the saturable nonlinearity $g(n)\propto n/(n_{\mathrm{s}}+n)$ ($n_{\mathrm{s}}>0$) specific to the superfluid-light experiment of Ref.~\cite{Eloy2021}, one of the flagship studies which have motivated the present work (see fifth paragraph of Appendix~\ref{App:A}), and in the case of the standard nonlinear Schr\"odinger potential $g(n)\propto n$ one has for dilute ultracold bosonic atoms or condensed exciton-polaritons (see second and fourth paragraphs of Appendix~\ref{App:A}). In what follows, we establish the hydrodynamic problem that the fluid must satisfy in order to flow in a superfluid fashion past the obstacle.

Replacing $\psi(x,t)$ with its density-phase representation (see above) into Eq.~\eqref{Eq:GNLSE} yields the following coupled hydrodynamic-like equations for $n(x,t)$ and $v(x,t)$:
\begin{equation}
\label{Eq:HydroEqs}
\begin{gathered}
n_{t}+(nv)_{x}=-\gamma n, \\
mv_{t}=-\bigg[\frac{mv^{2}}{2}+U(x)+g(n)-\frac{\hbar^{2}}{2m}\frac{\sqrt{n}_{xx}}{\sqrt{n}}\bigg]_{x}.
\end{gathered}
\end{equation}
The first of Eqs.~\eqref{Eq:HydroEqs} expresses the conservation of the number of particles when $\gamma=0$. The second one is formally identical to the Euler equation for the potential flow of an inviscid classical fluid with some additional potential (last term in the square brackets) of quantum origin \cite{Pitaevskii2016}. It is worth noting that the nonconservative term $-\gamma n$ in the first of Eqs.~\eqref{Eq:HydroEqs} makes the dynamics quite harsh to tackle analytically. To circumvent this drawback, we temporarily set $\gamma=0$ and postpone the analysis of the effect of particle losses to Sec.~\ref{Sec:EffectOfParticleLosses}.

\begin{figure}[t!]
\includegraphics[width=\linewidth]{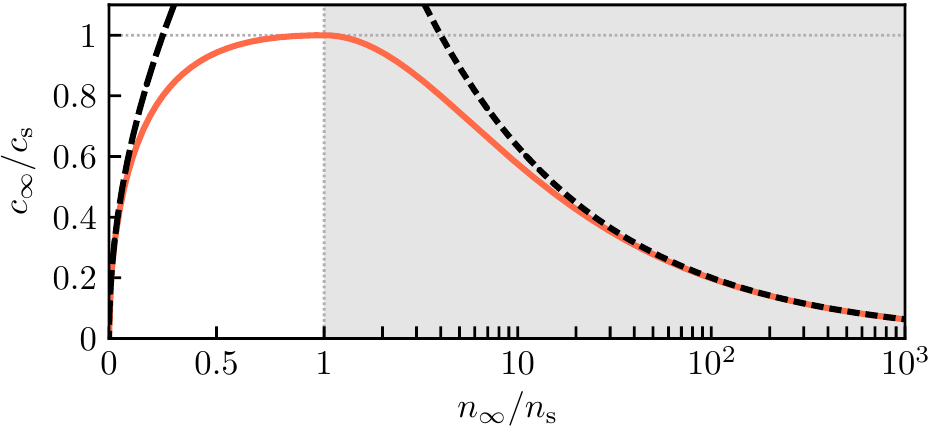}
\caption{The solid curve represents the Landau critical velocity for superfluidity $c_{\infty}$ (see main text) as a function of $n_{\infty}$ when $g(n)\propto n/(n_{\mathrm{s}}+n)$ (on the vertical axis, $c_{\mathrm{s}}=[g'(n_{\mathrm{s}})n_{\mathrm{s}}/m]^{1/2}$). The dashed (densely dashed) curve represents the $\sqrt{n_{\infty}}$ ($1/\sqrt{n_{\infty}}$) behavior of $c_{\infty}$ at low (high) $n_{\infty}$ \cite{NoteSoundVelocity}. In the shaded area, the horizontal axis is set in logarithmic scale to show the slow decreasing of $c_{\infty}$ at large $n_{\infty}$.}
\label{Fig:CriticalVelocitySF1D_Landau}
\end{figure}

In the absence of the obstacle, i.e., when $U(x)=0$, the simplest solutions of Eqs.~\eqref{Eq:HydroEqs} are homogeneous and stationary: $n(x,t)=n_{\infty}=\mathrm{const}$ and $v(x,t)=v_{\infty}=\mathrm{const}$, which describes a uniform and steady flow. In the comoving frame, the elementary excitations of such a system (typically stimulated by a weak external perturbation) obey the well-known \cite{Pitaevskii2016} Bogoliubov dispersion relation $\varepsilon(p)=[g'(n_{\infty})n_{\infty}p^{2}/m+p^{4}/(4m^{2})]^{1/2}$, from which we infer that the Landau critical velocity for superfluidity is the speed of sound $c_{\infty}=[g'(n_{\infty})n_{\infty}/m]^{1/2}$, slope of the linear branch of $\varepsilon(p)$ for $0<p\ll\hbar/\xi_{\infty}$, where $\xi_{\infty}=\hbar/(mc_{\infty})$ denotes the healing length. Figure~\ref{Fig:CriticalVelocitySF1D_Landau}
shows it against $n_{\infty}$ for $g(n)\propto n/(n_{\mathrm{s}}+n)$.

\begin{figure}[t!]
\includegraphics[width=\linewidth]{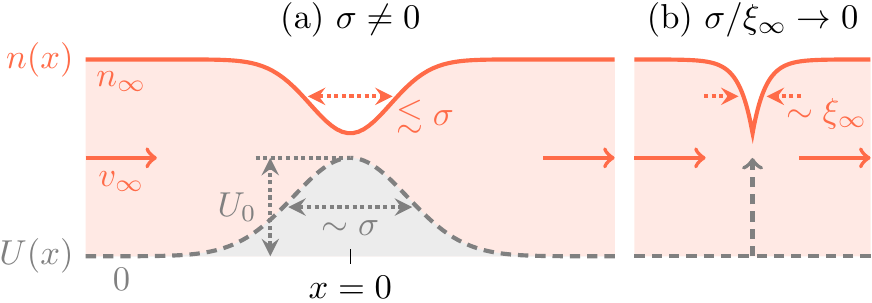}
\caption{(a) Schematic of the superfluid density (solid curve) when the obstacle potential (dashed curve) is repulsive and of typical width $\sigma\neq0$. The density dip in its vicinity is smooth and its typical width is $<\sigma$ or $\sim\sigma$ when $\sigma/\xi_{\infty}\to\infty$, where $\xi_{\infty}=1$ is the healing length of the unperturbed fluid. (b) Same as in panel~(a) but when $\sigma/\xi_{\infty}\to0$. In this limit, the obstacle potential can be approximated by a Dirac $\delta$ function at the location of which the density dip has a discontinuous derivative and a typical width $\sim\xi_{\infty}$. When the obstacle potential is attractive, it induces a density bump of analogous characteristics in its vicinity.}
\label{Fig:Sketch}
\end{figure}

In the presence of the obstacle instead, $n(x,t)$ and $v(x,t)$ are no longer homogeneous. We search for them in the stationary forms $n(x,t)=n(x)$ and $v(x,t)=v(x)$ and suppose that they smoothly approach $n_{\infty}$ and $v_{\infty}$, respectively, out of range of $U(x)$, i.e., as $|x|\gg\sigma$. These hypotheses are typical of a superfluid flow, steady and devoid of any hydrodynamic disturbance far away from the obstacle \cite{Pavloff2002, NoteAsymptotics}. With this, Eqs.~\eqref{Eq:HydroEqs} may be cast into a single equation for $n(x)$:
\begin{equation}
\label{Eq:HydroEqDensity}
\frac{\hbar^{2}}{2m}\frac{\sqrt{n}''}{\sqrt{n}}+\frac{mv_{\infty}^{2}}{2}\bigg(1-\frac{n_{\infty}^{2}}{n^{2}}\bigg)+g(n_{\infty})-g(n)=U(x),
\end{equation}
the solution of which determines the velocity according to $v(x)=v_{\infty}n_{\infty}/n(x)$. From now on, we simplify the notations by expressing all the densities, velocities, lengths, and energies in units of $n_{\infty}$, $c_{\infty}$, $\xi_{\infty}$, and $\mu_{\infty}^{\vphantom{2}}=mc_{\infty}^{2}$, respectively. In these new variables, Eq.~\eqref{Eq:HydroEqDensity} takes the dimensionless form
\begin{equation}
\label{Eq:HydroEqDensityBis}
\frac{1}{2}\frac{\sqrt{n}''}{\sqrt{n}}+\frac{v_{\infty}^{2}}{2}\bigg(1-\frac{1}{n^{2}}\bigg)+g(1)-g(n)=U(x),
\end{equation}
where $n(x)\to1$ as $|x|\gg\sigma$. Figure~\ref{Fig:Sketch}
gives a schematic picture of the solutions of Eq.~\eqref{Eq:HydroEqDensityBis} we are looking for. The superfluid-nonsuperfluid transition, driven by the injection flow velocity $v_{\infty}$ in Eq.~\eqref{Eq:HydroEqDensityBis}, occurs when such solutions no longer exist. This equivalently corresponds to the appearance of an energetic instability in the time-dependent Eqs.~\eqref{Eq:HydroEqs} or Eq.~\eqref{Eq:GNLSE} \cite{Pitaevskii2016}. Energy is in this case typically dissipated by the emission of solitons \cite{Hakim1997, Leboeuf2001, Pavloff2002}.

It is instructive to linearize Eq.~\eqref{Eq:HydroEqDensityBis} for $|x|\gg\sigma$, there where $U(x)$ is zero and $\delta n(x)=n(x)-1$ must be small. This gives $\delta n''-4(1-v_{\infty}^{2})\delta n=0$, which admits suitable, vanishing solutions only if $v_{\infty}<1$, i.e., if the incoming flow is subsonic. One formally recovers Landau's criterion for superfluidity but to conclude that the critical velocity is the asymptotic speed of sound would be very cavalier. Indeed, no small-amplitude hypothesis has been made on $U(x)$ while Landau's approach requires such a constraint to deal with elementary excitations. On the other hand, it is a safe bet that the critical velocity depends on the characteristics of $U(x)$ while the present approach totally neglects $U(x)$. Nonetheless, by telling one that superfluidity cannot occur at supersonic injection speeds $v_{\infty}$, it allows us to restrict ourselves to the regime $v_{\infty}<1$, which provides an upper bound for the actual critical velocity.

Superfluidity actually depends on both the sign and the shape of the obstacle potential. For repulsive potentials, we show that it exists only if $v_{\infty}$ is smaller than a specific speed $v_{\mathrm{c}}<1$ function of the parameters of $U(x)$ [and obviously of $g(n)$]. For attractive potentials instead, we demonstrate that it is the rule for any $v_{\infty}<1$. Analytical results are obtained in the narrow- and wide-obstacle limits in Secs.~\ref{Sec:NarrowObstacle} and \ref{Sec:WideObstacle}, respectively. The obstacle of arbitrary width is numerically treated in Sec.~\ref{Sec:ObstacleOfArbitraryWidth}.

\section{Narrow obstacle}
\label{Sec:NarrowObstacle}

When the typical range of the obstacle potential is much smaller than the healing length of the unperturbed fluid, i.e., in dimensionless units, when $\sigma\ll1$, it is possible to approximate $U(x)$ by $U(x)=U_{0}F(\sigma)\delta(x)$, where $F(\sigma)$ is the integral of $f(|x|/\sigma)$ over the whole real axis. For example, with $F(\sigma)=\sqrt{\pi}\sigma$, this fairly approximates $U(x)=U_{0}\exp(-x^{2}/\sigma^{2})$ at small width $\sigma$.

In this case and when $U_{0}>0$ (repulsive obstacle potential), we find that the critical velocity for superfluidity $v_{\mathrm{c}}$ is a function of $U_{0}$ and $\sigma$ smaller than $1$ implicitly given by [we define the antiderivative $G(n)=\int dn\,g(n)$]
\begin{align}
\notag
\sqrt{2}\bigg[&{-}\frac{v_{\mathrm{c}}^{2}}{2}\bigg(1-\frac{1}{n_{0,\mathrm{c}}}\bigg)^{2} \\
\label{Eq:VcNarrow}
&{-}\,g(1)+\frac{g(1)-G(1)+G(n_{0,\mathrm{c}})}{n_{0,\mathrm{c}}}\bigg]^{\frac{1}{2}}=U_{0}F(\sigma),
\end{align}
where $n_{0,\mathrm{c}}<1$, the density of the fluid at $x=0$ when $v_{\infty}=v_{\mathrm{c}}$, is solution of
\begin{equation}
\label{Eq:N0cNarrow}
\frac{n_{0,\mathrm{c}}}{1-n_{0,\mathrm{c}}}[g(1)-G(1)-g(n_{0,\mathrm{c}})n_{0,\mathrm{c}}+G(n_{0,\mathrm{c}})]=v_{\mathrm{c}}^{2}.
\end{equation}
The derivation of Eqs.~\eqref{Eq:VcNarrow} and \eqref{Eq:N0cNarrow} is carefully detailed in Appendix~\ref{App:B}.

In Fig.~\ref{Fig:CriticalVelocitySF1D_Narrow}, we plot $v_{\mathrm{c}}$ as a function of $U_{0}F(\sigma)$ for $g(n)=(1+n_{\mathrm{s}})^{2}n/[n_{\mathrm{s}}(n_{\mathrm{s}}+n)]$ with $n_{\mathrm{s}}\in\{0.1,1,10\}$ and for $g(n)=n$ [saturable potential $g(n)\propto n/(n_{\mathrm{s}}+n)$ and nonlinear Schr\"odinger potential $g(n)\propto n$, respectively, in units of $\mu_{\infty}$; see Sec.~\ref{Sec:SuperfluidHydrodynamics}]. A few comments are in order. First, one sees that $v_{\mathrm{c}}\to1$ as $U_{0}F(\sigma)\to0$. This is in agreement with what the Landau criterion for superfluidity forecasts in this case: In the presence of a weakly perturbing obstacle, the fluid dissipates energy only through the emission of Bogoliubov excitations, whose phase velocity has for minimum $c_{\infty}$, i.e., in dimensionless units, $1$ (see Sec.~\ref{Sec:SuperfluidHydrodynamics}). On the other hand, $v_{\mathrm{c}}$ decreases and tends to $0$ as $U_{0}F(\sigma)$ increases, which is physically expected too: The stronger the obstacle, the greater the resistance to it and the less likely superfluidity is to occur. Finally, it is interesting to note that the critical velocity for light superfluidity (red curves in Fig.~\ref{Fig:CriticalVelocitySF1D_Narrow}) majorizes the one associated to the standard nonlinearity $g(n)=n$ (blue curve). This gives credit to fluid-of-light experiments for studying superfluidity \cite{Vocke2016, Michel2018, Glorieux2018, Eloy2021}, which should be more robust against short-range perturbations than in experiments with bosonic atoms or exciton-polaritons (disregarding photon absorption which has the tendency to reduce the critical velocity; see Sec.~\ref{Sec:EffectOfParticleLosses}). The same behavior is observed for long-range obstacle potentials (see Fig.~\ref{Fig:CriticalVelocitySF1D_Wide} afterwards).

\begin{figure}[t!]
\includegraphics[width=\linewidth]{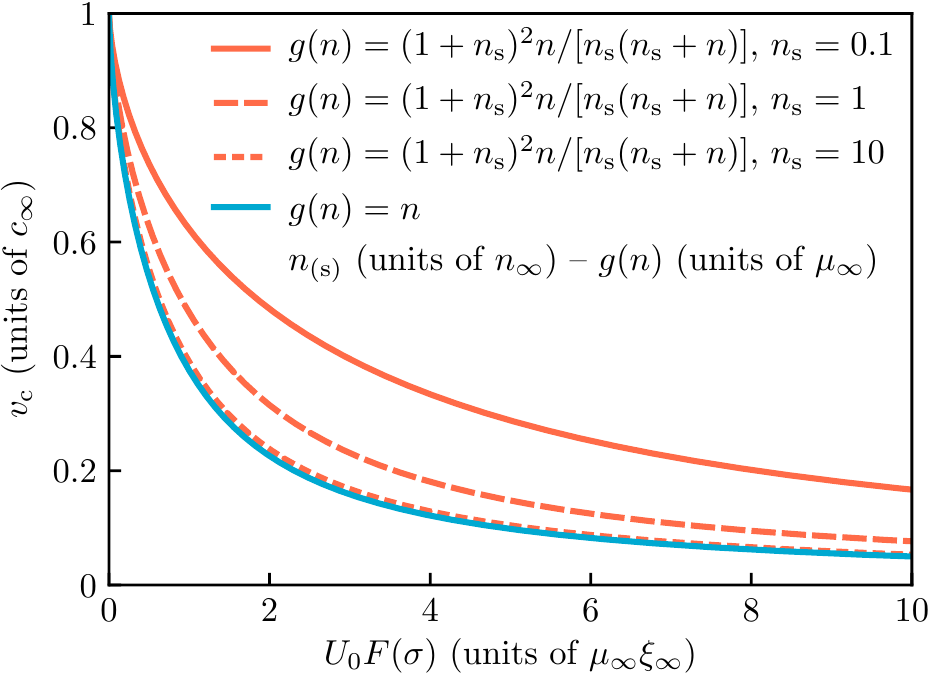}
\caption{Critical velocity for superfluidity $v_{\mathrm{c}}$ for a $\delta$-peaked obstacle potential of amplitude $U_{0}F(\sigma)>0$ [Eqs.~\eqref{Eq:VcNarrow} and \eqref{Eq:N0cNarrow}]. The two types of nonlinearity $g(n)$ considered here are indicated in the legend. They almost coincide and thus approximately give the same $v_{\mathrm{c}}$ when $n_{\mathrm{s}}$ is large, which can be seen on the graph where the densely dashed curve almost merges with the blue solid one.}
\label{Fig:CriticalVelocitySF1D_Narrow}
\end{figure}

When $U_{0}<0$ (attractive obstacle potential), we find that the flow can be superfluid for all $v_{\infty}<1$, hence $v_{\mathrm{c}}=1$ for all $U_{0}F(\sigma)$ (see Appendix~\ref{App:B}). This makes Landau's prediction valid even for large negative values of $U_{0}F(\sigma)$, as already found in Ref.~\cite{Pavloff2002} for $g(n)=n$. This can be understood within some local-density reformulation of the Landau criterion for superfluidity \cite{Hakim1997, Albert2008, Albert2009, Albert2010}, although it is not really appropriate in the narrow-obstacle limit $\sigma\ll1$. This approach is discussed at the start of Sec.~\ref{Sec:WideObstacle} where its condition of validity ($\sigma\gg1$) is satisfied.

\section{Wide obstacle}
\label{Sec:WideObstacle}

In the wide-obstacle limit $\sigma\gg1$, the gradients of $U(x)$ are small and so the fluid behaves almost as if it were uniform. This makes it possible to define a speed of sound $c(x)=\{g'[n(x)]n(x)\}^{1/2}$ [and then a healing length $\xi(x)=1/c(x)$, a ``chemical potential'' $\mu(x)=c^{2}(x)$, etc.] at any point $x$. This is the local-density approximation \cite{Pitaevskii2016}. According to Landau's criterion, the flow past the obstacle thus should be superfluid as long as the local flow velocity $v(x)=v_{\infty}/n(x)$ does not exceed the local sound velocity $c(x)$ \cite{Hakim1997, Albert2008, Albert2009, Albert2010}. This condition straightforwardly reformulates into $v_{\infty}<v_{\mathrm{c},\mathrm{LL}}$, where the critical velocity for superfluidity
\begin{equation}
\label{Eq:VcLL}
v_{\mathrm{c},\mathrm{LL}}=[g'(n_{\mathrm{min}}^{\vphantom{3}})n_{\mathrm{min}}^{3}]^{\frac{1}{2}}
\end{equation}
(``$\mathrm{LL}$'' stands for ``local Landau'') stems from the minimum $n_{\mathrm{min}}$ of the density $n(x)$. When the obstacle potential is repulsive, $x=0$ is the point where the particles are expelled the most (see Fig.~\ref{Fig:Sketch}). As a result, $n_{\mathrm{min}}=n(0)=n_{0}<1$ and $v_{\mathrm{c},\mathrm{LL}}=[g'(n_{0}^{\vphantom{3}})n_{0}^{3}]^{1/2}<1$. When the obstacle potential is attractive instead, $x=\pm\infty$ is the place where the particles mass together the less (see Fig.~\ref{Fig:Sketch} and remark at the end of its caption). Thus, $n_{\mathrm{min}}=n(\pm\infty)=1$ and $v_{\mathrm{c},\mathrm{LL}}=1$.

Although a priori limited to perturbative obstacles (it is based on Landau's criterion), the phenomenological approach detailed above is also accurate to describe the superfluid transition in the case of nonperturbative obstacles. We confirm this by determining the obstacle dependence of the critical velocity for superfluidity $v_{\mathrm{c}}$ from a rigorous multiple-scale treatment of the obstacle potential when $\sigma\gg1$. We take the latter in the form $U(x)=U_{0}[1+f''(0)x^{2}/(2\sigma^{2})]$, where the terms in square brackets correspond to the series expansion of $f(|x|/\sigma)$ to second order in $1/\sigma\ll1$. For example, with $f''(0)=-2$, this fairly approximates $f(|x|/\sigma)=\exp(-x^{2}/\sigma^{2})$ at large $\sigma$. In the following, we report and discuss the main results only, and refer the reader to Appendix~\ref{App:C} for the details of their derivation.

When $U_{0}>0$ (repulsive obstacle potential), we get the following formula for $v_{\mathrm{c}}$ as a function of $U_{0}$, $f''(0)<0$, and $\sigma$:
\begin{equation}
\label{Eq:VcWide}
v_{\mathrm{c}}=\bar{v}_{\mathrm{c}}+\frac{C}{2^{\frac{5}{3}}}\frac{[U_{0}|f''(0)|\bar{n}_{0,\mathrm{c}}]^{\frac{2}{3}}}{\bar{v}_{\mathrm{c}}(\frac{1}{\bar{n}_{0,\mathrm{c}}}-\bar{n}_{0,\mathrm{c}})[\frac{3\bar{v}_{\mathrm{c}}^{2}}{\bar{n}_{0,\mathrm{c}}^{3}}+g''(\bar{n}_{0,\mathrm{c}})\bar{n}_{0,\mathrm{c}}]^{\frac{1}{3}}}\frac{1}{\sigma^{\frac{4}{3}}},
\end{equation}
where $C\simeq1.466$. In Eq.~\eqref{Eq:VcWide}, $\bar{v}_{\mathrm{c}}$ denotes the critical velocity resulting from the first, $\sigma$-independent contribution to the series expansion of $U(x)$ written above. It is smaller than $1$ and its single dependence on $U_{0}$ is implicitly given by
\begin{equation}
\label{Eq:VcWide0}
\frac{\bar{v}_{\mathrm{c}}^{2}}{2}\bigg(1-\frac{1}{\bar{n}_{0,\mathrm{c}}^{2}}\bigg)+g(1)-g(\bar{n}_{0,\mathrm{c}})=U_{0},
\end{equation}
where $\bar{n}_{0,\mathrm{c}}<1$, the zeroth-order density of the fluid at $x=0$ when $v_{\infty}=\bar{v}_{\mathrm{c}}$, is solution of
\begin{equation}
\label{Eq:N0cWide0}
g'(\bar{n}_{0,\mathrm{c}}^{\vphantom{3}})\bar{n}_{0,\mathrm{c}}^{3}=\bar{v}_{\mathrm{c}}^{2}.
\end{equation}
As defined by Eq.~\eqref{Eq:N0cWide0}, $\bar{v}_{\mathrm{c}}$ coincides with the critical velocity $v_{\mathrm{c},\mathrm{LL}}$ deduced from the local Landau criterion for superfluidity [see discussion after Eq.~\eqref{Eq:VcLL}]. This removes the bias one could have on its accuracy to capture the superfluid transition in the case of arbitrary-high wide obstacles. Meanwhile, the nontrivial $1/\sigma^{4/3}$ correction to $\bar{v}_{\mathrm{c}}$ in Eq.~\eqref{Eq:VcWide} results from the second, $1/\sigma^{2}$ term in the series expansion of $U(x)$. One sees that it is positive, which is actually natural: At a fixed $U_{0}$, the narrower the obstacle, the easier it is for the fluid to cross it, thus favoring superfluidity.

\begin{figure}[t!]
\includegraphics[width=\linewidth]{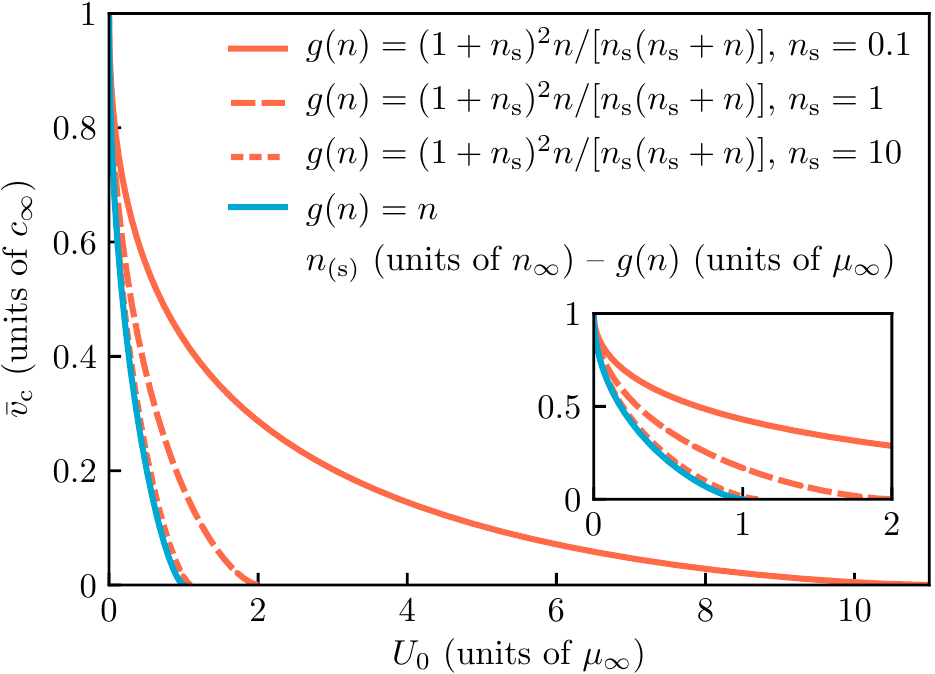}
\caption{Same as Fig.~\ref{Fig:CriticalVelocitySF1D_Narrow} but for a wide obstacle potential of maximum amplitude $U_{0}>0$, at zeroth order in its inverse range $1/\sigma$ [Eqs.~\eqref{Eq:VcWide0} and \eqref{Eq:N0cWide0}]. Each curve drops to zero at a specific $U_{0}=U_{0,\mathrm{max}}$ (see main text). In the case of the red solid (dashed, densely dashed, blue solid) curve, $U_{0,\mathrm{max}}=11$ ($2$, $1.1$, $1$). The inset magnifies the low-$U_{0}$ regime.}
\label{Fig:CriticalVelocitySF1D_Wide}
\end{figure}

In the following, we mainly focus on $\bar{v}_{\mathrm{c}}$ which we plot in Fig.~\ref{Fig:CriticalVelocitySF1D_Wide}
as a function of $U_{0}$ for the same parameters as the ones used in Fig.~\ref{Fig:CriticalVelocitySF1D_Narrow}. The $1/\sigma^{4/3}$ decreasing of $v_{\mathrm{c}}$ at large $\sigma$ will be shown in Sec.~\ref{Sec:ObstacleOfArbitraryWidth} where we numerically get all the $\sigma$ dependence of the critical velocity. Comparing Fig.~\ref{Fig:CriticalVelocitySF1D_Wide} to Fig.~\ref{Fig:CriticalVelocitySF1D_Narrow}, the same qualitative behavior is observed for $\bar{v}_{\mathrm{c}}$, except that the latter identically vanishes at $U_{0}=U_{0,\mathrm{max}}=g(1)-g(\bar{n}_{0,\mathrm{c}}|_{\bar{v}_{\mathrm{c}}=0}=0)$ [see Eqs.~\eqref{Eq:VcWide0} and \eqref{Eq:N0cWide0}]. For example, $U_{0,\mathrm{max}}=(1+n_{\mathrm{s}})/n_{\mathrm{s}}$ when $g(n)=(1+n_{\mathrm{s}})^{2}n/[n_{\mathrm{s}}(n_{\mathrm{s}}+n)]$ and $U_{0,\mathrm{max}}=1$ when $g(n)=n$ [more generally, $U_{0,\mathrm{max}}=1/\nu$ when $g(n)=n^{\nu}/\nu$, i.e., $g(n)\propto n^{\nu}$ in units of $\mu_{\infty}$; see second and third paragraphs of Appendix~\ref{App:A} for examples with $\nu\neq1$]. This abrupt cancelation of $\bar{v}_{\mathrm{c}}$ can be explained by simple words: The fact that $\bar{n}_{0,\mathrm{c}}=0$ when $U_{0}=U_{0,\mathrm{max}}$ means that the fluid is for such an obstacle fragmented into two disconnected parts on either side of $x=0$; superfluidity is impossible in this situation, hence $\bar{v}_{\mathrm{c}}=0$.

In reality, tunneling through the potential barrier always occurs when $U_{0}>U_{0,\mathrm{max}}$, but this cannot be captured within our multiple-scale analysis where the derivatives of the fluid density are assumed to scale as positive powers of $1/\sigma\ll1$, and even less at the here-discussed zeroth order where they are identically null. Indeed, the greater the density depletion in the vicinity of $x=0$, the less valid such a small-gradient hypothesis in this region of space is. When $U_{0}$ is very large, the obstacle actually behaves as the insulator layer of a Josephson junction crossed by a (conserved) supercurrent density~$j=n(x)v(x)=v_{\infty}=v_{\mathrm{c}}\sin(\theta_{\mathrm{R}}-\theta_{\mathrm{L}})$ ($\theta_{\mathrm{R}}-\theta_{\mathrm{L}}$ is the phase difference between the right and the left of the density depletion) whose amplitude $v_{\mathrm{c}}$, which naturally defines the critical velocity for superfluidity, is certainly very small but not zero (see Refs.~\cite{Sols1994, Zapata1996, TarasSemchuk1999, Danshita2006, Danshita2007, Watanabe2009} which mostly focus on $\delta$-shaped obstacles). In the limit $\sigma\gg1$, a WKB treatment of Eq.~\eqref{Eq:GNLSE} can be performed to get this $v_{\mathrm{c}}$ at very large $U_{0}$. This is done in Appendix~\ref{App:D} where $v_{\mathrm{c}}$ is found to be given by
\begin{equation}
\label{Eq:VcJosephson}
v_{\mathrm{c}}=A\exp\!\bigg\{{-}\sqrt{2}\int_{-x_{\mathrm{cl}}}^{x_{\mathrm{cl}}}dx\,[U(x)-\mu]^{\frac{1}{2}}\bigg\},
\end{equation}
where $\mu$ is the stationary-state energy of the quantum fluid and $\pm x_{\mathrm{cl}}$ are its classical turning points, solutions of $U(\pm x_{\mathrm{cl}})=\mu$. The amplitude $A$ is a function of the obstacle parameters dominated by the exponential above. It is proportional to $|U'(\pm x_{\mathrm{cl}})|^{(2+\nu)/(3\nu)}$ when $g(n)=n^{\nu}/\nu$ (see Appendix~\ref{App:D}).

When $U_{0}<0$ (attractive obstacle potential), we find, as in the case of the attractive $\delta$ peak, that $v_{\mathrm{c}}=1$ for all $U_{0}$, in accordance with the local Landau criterion for superfluidity [see discussion after Eq.~\eqref{Eq:VcLL}]. This result is analytically provable at zeroth order in $1/\sigma$ (see Appendix~\ref{App:C}). We numerically checked that it persists beyond---and even well beyond---zeroth order for the realistic obstacle potential $U(x)=U_{0}\exp(-x^{2}/\sigma^{2})$ and the different self-interaction potentials $g(n)$ given in Appendix~\ref{App:A}.

\section{Obstacle of arbitrary width}
\label{Sec:ObstacleOfArbitraryWidth}

Now that we understand the two extreme cases of a narrow ($\sigma\ll1$) and a wide ($\sigma\gg1$) obstacle, we move to the generic situation of a localized $U(x)=U_{0}f(|x|/\sigma)$ of arbitrary range, restricting to the case $U_{0}>0$ for which the critical velocity for superfluidity is not trivial (see last paragraphs of Secs.~\ref{Sec:NarrowObstacle} and \ref{Sec:WideObstacle}). Unfortunately, it is not possible to analytically solve Eq.~\eqref{Eq:HydroEqDensityBis} for any $\sigma$ and we thus have to resort to a numerical integration of it. In the following, we quite generically take $f(|x|/\sigma)=\exp(-x^{2}/\sigma^{2})$. Again, the separatrix between the superfluid and nonsuperfluid regimes is given by the threshold of disappearance of a subsonic stationary density profile smoothly approaching unity at infinity. This criterion, applied to the time-independent Eq.~\eqref{Eq:HydroEqDensityBis} we focus on, actually corresponds to the appearence of an energetic instability in the time-dependent Eqs.~\eqref{Eq:HydroEqs} or Eq.~\eqref{Eq:GNLSE} \cite{Pitaevskii2016}.

Numerically solving Eq.~\eqref{Eq:HydroEqDensityBis} with boundary conditions at infinity is actually cumbersome. For instance, a shooting method could be applied to this problem but eventually happens to be poorly efficient. We chose instead to use a method inspired by out-of-equilibrium statistical physics, which we sketch the contours right after. If the superfluid solution exists, it should be an attractor of the following fictitious dynamical system:
\begin{equation}
\label{Eq:FictitiousDynamics}
n_{\tau}=\frac{1}{2}\frac{\sqrt{n}_{xx}}{\sqrt{n}}+\frac{v_{\infty}^{2}}{2}\bigg(1-\frac{1}{n^{2}}\bigg)+g(1)-g(n)-U(x).
\end{equation}
We start the integration with a solution which has the proper asymptotic properties---typically the very simple homogeneous density $n(x;\tau=0)=1$---and let it evolve in the fictitious time $\tau$ according to the equation of motion \eqref{Eq:FictitiousDynamics}. After some time, the solution $n(x;\tau)$ converges to the attractor if the latter exists. If not, this means that the system is not superfluid.

\begin{figure}[t!]
\includegraphics[width=\linewidth]{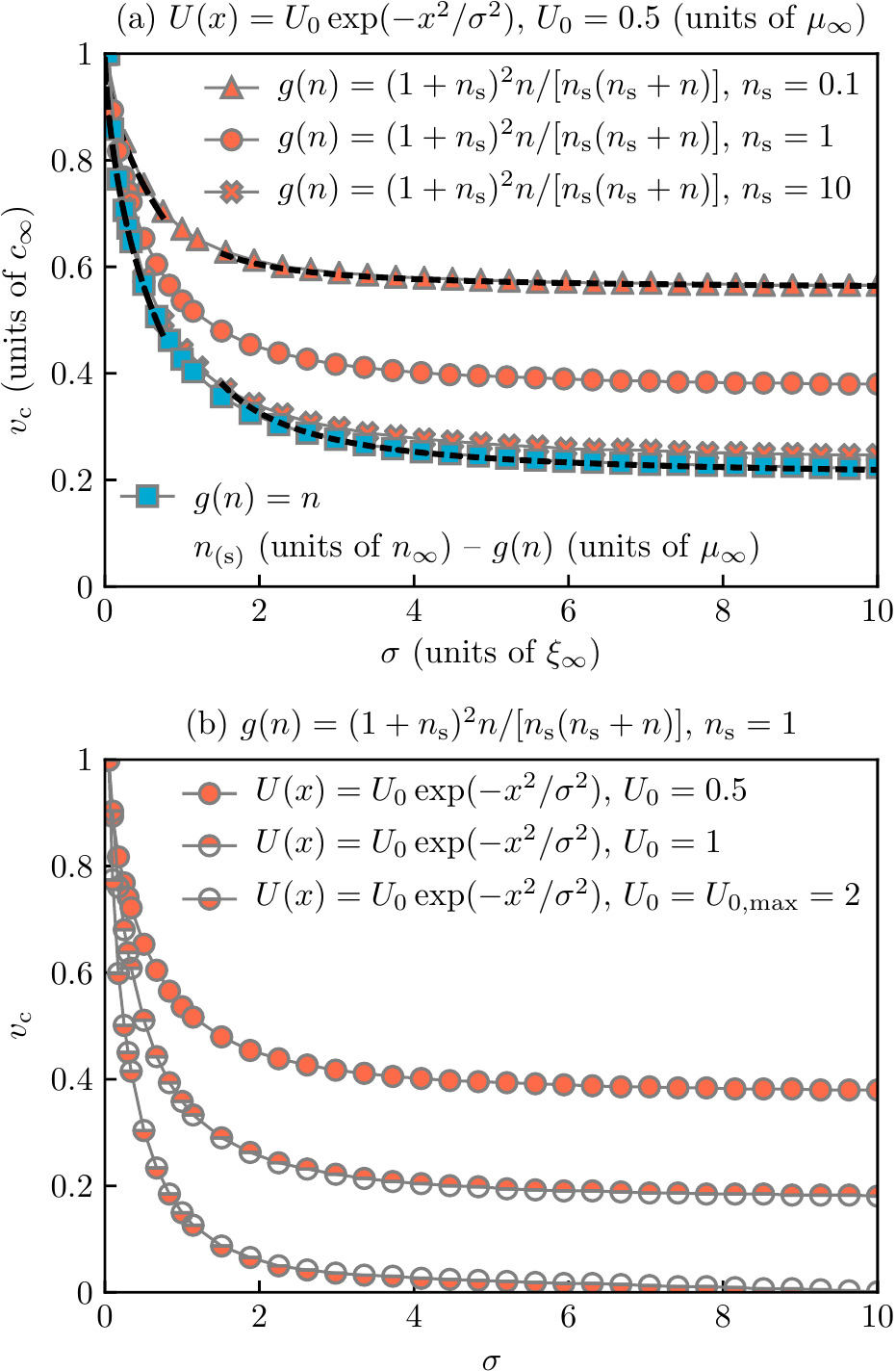}
\caption{Critical velocity for superfluidity $v_{\mathrm{c}}$ as a function of the typical range $\sigma$ of the obstacle potential $U(x)=U_{0}\exp(-x^{2}/\sigma^{2})$, supposed to be repulsive ($U_{0}>0$). It is determined from the numerical integration of the fictitious time-dependent problem \eqref{Eq:FictitiousDynamics}: The critical frontier $v_{\infty}=v_{\mathrm{c}}$ is crossed from the moment when the algorithm is no longer able to converge towards an attractor with the correct boundary conditions at infinity. On both plots, the numerical points are artificially joined by solid segments to guide the eye. In panel~(a), $v_{\mathrm{c}}$ is plotted at a fixed $U_{0}$ and for four different $g(n)$'s (see title and legend). The two types of $g(n)$ considered here almost coincide and thus approximately give the same $v_{\mathrm{c}}$ when $n_{\mathrm{s}}$ is large, which can be seen on the graph where the crossed data almost coincide with the squared ones. The asymptotic results \eqref{Eq:VcNarrow} and \eqref{Eq:N0cNarrow} ($\sigma\ll1$, dashed curves) and \eqref{Eq:VcWide}--\eqref{Eq:N0cWide0} ($\sigma\gg1$, densely dashed curves) are in perfect agreement with numerics. For the sake of lisibility, they are only shown on top of the triangled and squared data. In panel~(b), $v_{\mathrm{c}}$ is plotted at a fixed $g(n)$ and for three different $U_{0}$'s (see title and legend). When $U_{0}=U_{0,\mathrm{max}}$, one sees that $v_{\mathrm{c}}$ drops to zero at large $\sigma$, in conformity with the approach yielding Eqs.~\eqref{Eq:VcWide}--\eqref{Eq:N0cWide0}.}
\label{Fig:CriticalVelocitySF1D_Numerics}
\end{figure}

This relaxation method is rather simple, efficient, and accurate, as can be seen in Fig.~\ref{Fig:CriticalVelocitySF1D_Numerics}.
Figure~\ref{Fig:CriticalVelocitySF1D_Numerics}(a) displays the critical velocity for superfluidity $v_{\mathrm{c}}$ as a function of $\sigma$ for $U_{0}=0.5$, $g(n)=(1+n_{\mathrm{s}})^{2}n/[n_{\mathrm{s}}(n_{\mathrm{s}}+n)]$ with $n_{\mathrm{s}}\in\{0.1,1,10\}$, and $g(n)=n$. While the analytical results given by Eqs.~\eqref{Eq:VcNarrow} and \eqref{Eq:N0cNarrow} and by Eqs.~\eqref{Eq:VcWide}--\eqref{Eq:N0cWide0} are perfectly recovered at respectively small and large $\sigma$, the critical velocity is shown to vary monotonically from one limit to another. Figure~\ref{Fig:CriticalVelocitySF1D_Numerics}(b) displays the result for $g(n)=(1+n_{\mathrm{s}})^{2}n/[n_{\mathrm{s}}(n_{\mathrm{s}}+n)]$ with $n_{\mathrm{s}}=1$ and $U_{0}\in\{0.5,1,2\}$, the latter value of $U_{0}$ corresponding to the maximum obstacle strength $U_{0,\mathrm{max}}=(1+n_{\mathrm{s}})/n_{\mathrm{s}}$ above which superfluidity is lost in the small-gradient approach yielding Eqs.~\eqref{Eq:VcWide}--\eqref{Eq:N0cWide0}.

Although the method is illustrated in Fig.~\ref{Fig:CriticalVelocitySF1D_Numerics} for two specific $g(n)$'s, we have checked that the agreement is similar for other nonlinearities. The only drawback of the method is the determination of $v_{\mathrm{c}}$ for $U_{0}>U_{0,\mathrm{max}}$ where the physics strongly depends on the gradients of the density and is actually ruled by the Josephson effect (see second-last paragraph of Sec.~\ref{Sec:WideObstacle}). It is however possible to improve the relaxation algorithm by letting $U(x)$ depend on the fictitious time $\tau$ in Eq.~\eqref{Eq:FictitiousDynamics} and by slowly increasing its amplitude from $U_{0}=U_{0,\mathrm{max}}$. This yields exponentially small critical velocities of the type \eqref{Eq:VcJosephson}.

\section{Effect of particle losses}
\label{Sec:EffectOfParticleLosses}

So far, we have considered the dynamics to be conservative by neglecting particle losses, described by the $\gamma$ term in Eq.~\eqref{Eq:GNLSE}. However, in actual experiments, this term might nonnegligibly contribute to the dynamics. For example, in the fluid-of-light experiment of Ref.~\cite{Eloy2021}, photon absorption is estimated to be about $30\%$ of the input light intensity after propagation through the crystal. Losses are thus significant in this study but superfluid features are not thereby destroyed, as previously observed in exciton-polariton quantum fluids, yet very nonequilibrium by nature \cite{Carusotto2013}. The authors of Ref.~\cite{Eloy2021} justify their observations by invoking the principle of adiabatic evolution, also used in Refs.~\cite{Larre2017, Fontaine2020} and which we explain in the following lines.

If losses manifest on a time scale $1/\gamma$ much longer than any other hydrodynamic time [$\xi_{\infty}/c_{\infty}=\hbar/\mu_{\infty}$, the time needed to generate an excitation, etc.], it is reasonable to assume that the full dynamics of the fluid must adiabatically follow the time variations of its unperturbed density, which in our case decays exponentially as $n_{\infty}(t)=n_{\infty}(0)\exp(-\gamma t)$ since losses are supposed to be linear. In the context of Ref.~\cite{Eloy2021}, with $t=z$ (see fifth paragraph of Appendix~\ref{App:A}), this is nothing but the Beer-Lambert attenuation law for the optical intensity. This somehow amounts to treating the fluid as if it were in a stationary equilibrium state at each time $t$, decribed by Eq.~\eqref{Eq:HydroEqDensityBis} where the densities, velocities, lengths, and energies are respectively expressed in units of the instantaneous density $n_{\infty}(t)$, sound speed $c_{\infty}(t)=\{g'[n_{\infty}(t)]n_{\infty}(t)/m\}^{1/2}$, healing length $\xi_{\infty}(t)=\hbar/[mc_{\infty}(t)]$, and ``chemical potential'' $\mu_{\infty}^{\vphantom{2}}(t)=mc_{\infty}^{2}(t)$.

The purpose of this section is to provide a convincing numerical proof of this argument, focusing on the central object of the paper, namely the critical velocity for superfluidity. If correct, this adiabatic-evolution approximation should predict a critical velocity $\tilde{v}_{\mathrm{c}}(t)$ not too far from the critical velocity $v_{\mathrm{c}}(t)$ deduced from the numerical integration of the time-dependent Eq.~\eqref{Eq:GNLSE} with $\gamma\neq0$, if properly definable. We show that it is the case, and even that the agreement is extremely good for losses of the type of Ref.~\cite{Eloy2021}.

\begin{figure}[t!]
\includegraphics[width=\linewidth]{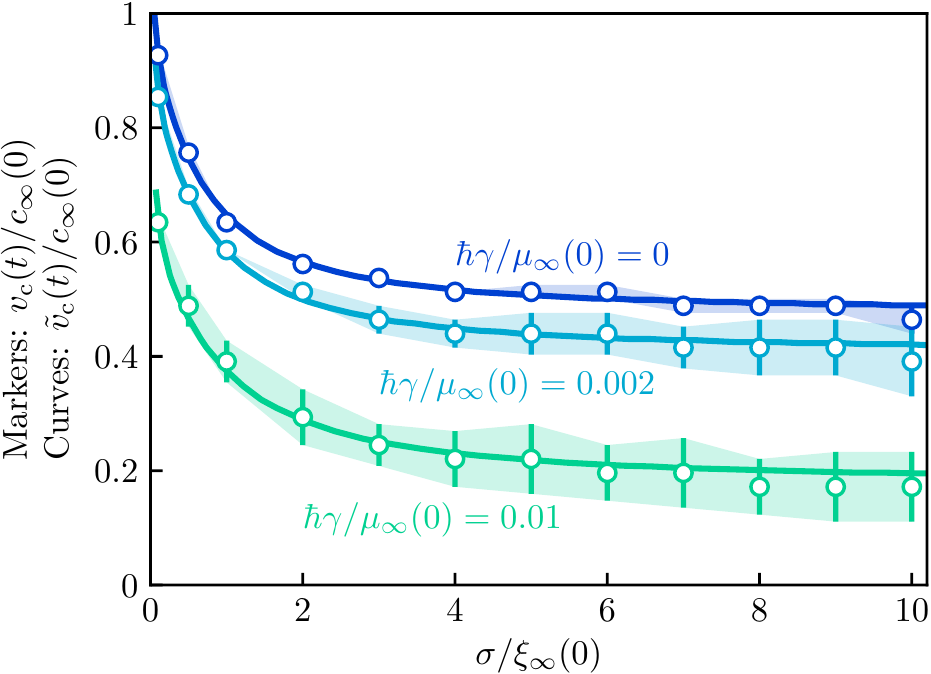}
\caption{Same as Fig.~\ref{Fig:CriticalVelocitySF1D_Numerics} for $g(n)\propto n$ but in the presence of particle losses [$\gamma$ term in Eq.~\eqref{Eq:GNLSE}]. The propagation time is here fixed to $\mu_{\infty}(0)t/\hbar=75$, the obstacle height to $U_{0}/\mu_{\infty}(0)=0.2$, and the loss rate to the three dimensionless values $\hbar\gamma/\mu_{\infty}(0)$ indicated on the plot. The markers correspond to the critical velocity for superfluidity $v_{\mathrm{c}}(t)$ obtained by numerically integrating Eq.~\eqref{Eq:GNLSE}. They are tainted with vertical error bars corresponding to the intervals of values of $v_{\infty}$ over which the drag force experienced by the obstacle (see main text) increases from $2\%$ to $8\%$ of its value at $v_{\infty}=c_{\infty}(t)>v_{\mathrm{c}}(t)$. Given the losses considered here, this provides correct confidence intervals for the superfluid-nonsuperfluid transition. The shaded regions guide the eye from one error bar to another. The curves correspond to the critical velocity for superfluidity $\tilde{v}_{\mathrm{c}}(t)$ deduced from the numerical integration of Eq.~\eqref{Eq:FictitiousDynamics} in our adiabatic-evolution approximation. In fact, they correspond to the solid segments used in Fig.~\ref{Fig:CriticalVelocitySF1D_Numerics} to guide the eye from one numerical point to another. The agreement between $v_{\mathrm{c}}(t)$ and $\tilde{v}_{\mathrm{c}}(t)$ is excellent, which validates the adiabatic hypothesis.}
\label{Fig:CriticalVelocitySF1D_Losses}
\end{figure}

For concreteness, we numerically integrate Eq.~\eqref{Eq:GNLSE} with $U(x,t)=U_{0}\exp[-(x+v_{\infty}t)^{2}/\sigma^{2}]$ (by a trivial change of reference frame, this is equivalent to the flow configuration studied so far). This brings two important informations on the superfluid-nonsuperfluid transition. First, the latter is not as sharp as in the conservative case. Losses have the tendency to smooth out the transition, as already observed in exciton-polariton condensates \cite{Carusotto2013} and which in principle undermines the very concept of critical velocity. Nonetheless, for losses of the type of Ref.~\cite{Eloy2021}, the drag force experienced by the obstacle, naturally defined as the average value of $U_{x}(x,t)$ over the wavefunction $\psi(x,t)$ \cite{Pavloff2002}, still increases very fast from (almost) zero on a narrow window of values of $v_{\infty}$ \cite{Wouters2010, Berceanu2012, Larre2012}, which makes it possible to identify an effective critical velocity for superfluidity $v_{\mathrm{c}}(t)$ without too many errors (details are provided in the caption of Fig.~\ref{Fig:CriticalVelocitySF1D_Losses}).
Second, this $v_{\mathrm{c}}(t)$ is smaller than the critical velocity in the absence of losses. This can be understood within the adiabatic-evolution hypothesis sketched above (and which we show the validity in Fig.~\ref{Fig:CriticalVelocitySF1D_Losses}). Indeed, a lower density [$n_{\infty}(t)<n_{\infty}(0)$] implies a smaller sound velocity [$c_{\infty}(t)<c_{\infty}(0)$] and therefore not only the Landau critical velocity, which gives a rough estimate of the actual critical velocity, is reduced but also the dimensionless amplitude of the obstacle potential is increased [$U_{0}/\mu_{\infty}(t)>U_{0}/\mu_{\infty}(0)$], making the superfluid window narrower.

Figure~\ref{Fig:CriticalVelocitySF1D_Losses} displays $v_{\mathrm{c}}(t)/c_{\infty}(0)$ as a function of $\sigma/\xi_{\infty}(0)$ at the dimensionless time $\mu_{\infty}(0)t/\hbar=75$ for $U_{0}/\mu_{\infty}(0)=0.2$ and $\hbar\gamma/\mu_{\infty}(0)\in\{0,0.002,0.01\}$. The nonlinearity $g(n)$ considered here is proportional to the density $n$ \cite{NoteNonlinearity}. At the time $t$ and in between the two latter values of $\gamma$ given above, losses range from about $14\%$ to $53\%$ of the initial unperturbed density, interval into which the photon absorption measured in the experiment of Ref.~\cite{Eloy2021} falls down. Most importantly, we compare on the plot $v_{\mathrm{c}}(t)$ with its adiabatic counterpart $\tilde{v}_{\mathrm{c}}(t)$. The latter is deduced from the numerical integration of the fictitious problem \eqref{Eq:FictitiousDynamics} where the dimensionless densities, velocities, lengths, and energies are such as defined in the second paragraph of the present section. For the sake of coherence, the algorithm behind Eq.~\eqref{Eq:FictitiousDynamics} thus must be run for $U_{0}/\mu_{\infty}(t)=\exp(\gamma t)U_{0}/\mu_{\infty}(0)\simeq0.232$ and $\sigma/\xi_{\infty}(t)=\exp(-\gamma t/2)\sigma/\xi_{\infty}(0)\simeq0.928\sigma/\xi_{\infty}(0)$ when we choose $\gamma=0.002$ in Eq.~\eqref{Eq:GNLSE}, and for $U_{0}/\mu_{\infty}(t)\simeq0.423$ and $\sigma/\xi_{\infty}(t)\simeq0.687\sigma/\xi_{\infty}(0)$ when $\gamma=0.01$. The agreement is excellent, which gives much credit to the adiabatic approach for explaning absorption features observed in typical experiments on light superfluidity \cite{Vocke2016, Michel2018, Glorieux2018, Eloy2021}, for example.

\section{Conclusion and perspectives}
\label{Sec:ConclusionAndPerspectives}

In this paper, we have studied the critical velocity for superfluidity of a generic quantum fluid flowing past a smooth localized obstacle of arbitrary amplitude and width in the 1D mean-field regime. Our model of superflow relies on a generalization of the 1D nonlinear Schr\"odinger equation to any local self-interaction potential increasing with the fluid density and to the eventuality of particle losses in the system, which we have treated within an accurate adiabatic treatment \cite{Larre2017, Fontaine2020, Eloy2021} of the fluid dynamics. Importantly, we have computed this critical velocity beyond the Landau criterion for superfluidity, which is valid only for weak-amplitude obstacles. We have derived nonperturbative exact expressions for the latter in the narrow- and wide-obstacle limits (following Refs.~\cite{Hakim1997, Leboeuf2001, Pavloff2002}), numerically computed it for any obstacle width using an original relaxation algorithm for the stationary problem, and quantitatively analyzed the residual Josephson-type tunneling occuring through a very high and slowly varying barrier (following Ref.~\cite{Dalfovo1996}). Together with previously mentioned existing studies \cite{Sols1994, Zapata1996, Hakim1997, TarasSemchuk1999, Leboeuf2001, Pavloff2002, Abid2003, Danshita2006, Danshita2007, Albert2008, Albert2009, Watanabe2009, Albert2010, Larre2012, Larre2015a, Arabahmadi2021a, Arabahmadi2021b}, this now gives a very broad picture of the critical velocity for superfluidity in the 1D mean-field regime.

Many experimental systems are well described by the generalized nonlinear Schr\"odinger equation \eqref{Eq:GNLSE} and our predictions can be tested and used in several ways. Concrete examples are reviewed in detail in Appendix~\ref{App:A}. These include ultracold atomic Bose-Einstein condensates and Fermi superfluids \cite{Pitaevskii2016} tightly confined along a line using highly anisotropic traps, exciton-polariton condensates \cite{Carusotto2013} in wire-shaped semiconductor optical microcavities, and fluids of light \cite{Leboeuf2010, Larre2015b, Vocke2016, Santic2018, Michel2018, Cherroret2018, Glorieux2018, Eloy2021} in effective 1D configurations.

In the condensates, the obstacle is usually created from the interaction between the vapor and an external laser \cite{Raman1999, Onofrio2000, Raman2001, Engels2007, Neely2010, Dries2010, Desbuquois2012}. The width of the obstacle (related to the waist of the laser) and the healing length can be controlled independently and chosen to be of the same order or different by at least one order of magnitude, so does its relative amplitude (related to the intensity of the laser) to the chemical potential. The relative motion between the fluid and the obstacle can be monitored either by sweeping the laser through the condensate \cite{Engels2007} or by exciting dipole oscillations in the gas \cite{Dries2010}. Then, the superfluid-nonsuperfluid transition can be probed by measuring the threshold of appearance of excitations or of the damping of the system's dipole oscillations. Similar experiments can be conducted with fermionic superfluids \cite{Miller2007}.

In experiments with polaritons \cite{Amo2009, Lerario2017}, the flow of the fluid around some structural defect in the cavity is usually generated by means of a resonant pump in nonzero incidence with the sample. The superfluid-nonsuperfluid transition, characterized by the appearance of disturbances in the fluid density, is revealed in both the near- and far-field images of the light outgoing from the cavity. In related experiments based on coherent light propagating in saturable nonlinear optical media \cite{Michel2018, Eloy2021}, often discussed in this work, superfluidity manifests in the paraxial diffraction of a laser in grazing incidence on an elongated refractive-index defect induced by a second laser. The incidence angle determines the flow velocity and the parameters of the defect, the ones of the obstacle. The superfluid-nonsuperfluid transition can be extracted from the measurement of an optical analog of the drag force experienced by the obstacle \cite{Michel2018} or from the real-space imaging of the transmitted intensity, which is defect-disturbed or not \cite{Eloy2021}.

As a perspective to the present theoretical work, it is natural to raise the question of the critical velocity for superfluidity in the 2 or 3D mean-field regime. As mentioned in the introduction, most of our knowledge is restricted to wide impenetrable obstacles in the standard nonlinear Schr\"odinger framework [nonlinearity of the form $g(n)\propto n$] \cite{Frisch1992, Josserand1997, Josserand1999, Berloff2000, Huepe2000, Stiessberger2000, Rica2001, Sasaki2010, Pinsker2014, Weimer2015, Singh2016, Singh2017, Pigeon2021}. A systematic study of the critical velocity, along the lines of the present work, is therefore highly desired: First in the presence of a single but penetrable impurity for various types of nonlinearity, but also in the presence of several defects or even a random environment. In the latter case, the critical velocity becomes a random variable \cite{Albert2010} whose statistical properties are so far completely unknown in 2 and 3D.

In any dimension, increasing the velocity of the superfluid brings the system into a time-dependent turbulent regime. This regime is called turbulent in the sense that maintaining a constant flux continuously creates excitations such as quantized vortices. These excitations interact with each other and may lead to quantum turbulence \cite{Barenghi2014}. However, at even higher velocity, the proliferation of excitations is strongly reduced due to the fact that all energy scales become negligible compared to the kinetic energy (provided the obstacle is penetrable). Concretely, nonlinear Sch\"odinger equations of the type \eqref{Eq:GNLSE} predict the existence of stationary solutions above some supersonic separatrix which should depend on the fluid and obstacle characteristics. This stationary regime is reminiscent of the noninteracting case described by linear-wave physics with however nonlinear modifications of the wave patterns. The equation of this separatrix is only known in the case of the 1D nonlinear Schr\"odinger equation \cite{Leboeuf2001, Leszczyszyn2009, Kamchatnov2012} (see also Ref.~\cite{Larre2012} for a numerical estimate of the latter in the case of a 1D incoherently pumped exciton-polariton condensate). It is therefore needed to develop a more general theory for it [arbitrary dimension, arbitrary $g(n)$], all the more so as experimental data are available in 2D for a saturable nonlinearity \cite{Eloy2021}. In 1D specifically, it could in principle be obtained from Eq.~\eqref{Eq:HydroEqDensityBis} with $v_{\infty}=v(x\to+\infty)>1$ by means of procedures similar to the ones used in this paper. We leave this for future work.

Another important perspective of this work is to understand the effect of quantum fluctuations on the superfluid-nonsuperfluid transition. Indeed, nonlinear Schr\"odinger-type equations only describe the dynamics of the mean-field state of superfluids, if it exists. If not, as it is the case in reduced dimensions on length scales smaller than the phase coherence length, in the presence of large interactions, or for small particle numbers, fully quantum descriptions are needed. Theoretical works \cite{Astrakharchik2004, Roberts2005, Roberts2006, Pomeau2008, Sykes2009, Cherny2012, Schenke2012, Lang2015, Reichert2019} already investigated superflows past heavy impurities beyond the mean-field regime, in the weak- and strong-interaction limits, but yet only a very few (mainly Ref.~\cite{Sykes2009}) went beyond a perturbative treatment of the obstacle as we do (at the mean-field level) in the present paper.

\begin{acknowledgments}
We gratefully acknowledge M. Bellec and C. Michel for inspiring exchanges on light superfluidity, M. Bellec again, F. H\'ebert, and N. Pavloff for their careful critical reading of the manuscript, and E. Arabahmadi, O. Lychkovskiy, V. Pastukhov, and V. P. Singh for sharing their works on quantum transport with us. This work has benefited from the financial support of the French Agence Nationale de la Recherche (ANR) under Grants No.~ANR-21-CE30-0008-01 (project STLight: Superfluid and Turbulent Light in Complex Media) and No.~ANR-21-CE47-0009 (project Quantum-SOPHA: Quantum Simulators for One-Dimensional Systems with Photons and Atoms).
\end{acknowledgments}

\appendix

\section{Superfluids described by Eq.~\texorpdfstring{\eqref{Eq:GNLSE}}{Lg}}
\label{App:A}

The generalized nonlinear Schr\"odinger equation~\eqref{Eq:GNLSE} rules the mean-field dynamics of various 1D superfluid systems which we give four examples in this appendix.

For instance, with $\gamma=0$, Eq.~\eqref{Eq:GNLSE} accurately describes the evolution of a zero-temperature dilute Bose-Einstein condensate of weakly repulsive identical atoms of mass $m$ in a harmonic trap of angular frequencies $\omega_{x}\ll\omega_{y}=\omega_{z}=\omega_{\perp}$, i.e., highly asymmetric to make the dynamics of the condensate quasi-1D along the $x$ axis \cite{Pitaevskii2016}. In this context, the obstacle potential $U(x)$ can be realized by crossing the cigar-shaped atomic cloud, elongated in the $x$ direction, with a detuned laser beam of $y$ (or $z$) axis and whose waist in the $z$ (or $y$) direction is larger than the typical transverse size of the condensate \cite{Engels2007}. On the other hand, the self-interaction potential $g(n)$, functional of the longitudinal density $n(x,t)$ of the condensate, is given by $g(n)=2\hbar\omega_{\perp}na_{\mathrm{s}}$ when $na_{\mathrm{s}}\ll1$ and by $g(n)=2\hbar\omega_{\perp}(na_{\mathrm{s}})^{1/2}$ when $na_{\mathrm{s}}\gg1$ \cite{Menotti2002, Gerbier2004}, where $a_{\mathrm{s}}$ is the $\mathrm{s}$-wave scattering length of the two-body interaction potential. Finally, the flow of the Bose fluid in the positive-$x$ direction can be simulated by sweeping the laser sheet creating the obstacle in the negative-$x$ direction \cite{Engels2007} and then by looking at the whole system in the reference frame of the obstacle.

Always with $\gamma=0$, Eq.~\eqref{Eq:GNLSE} is also well suited to describe the evolution of a zero-temperature dilute Fermi gas of interacting spin-$1/2$ atoms of mass $m/2$ in the same trapping configuration as the one depicted right above \cite{Pitaevskii2016}. The obstacle potential $U(x)$ experienced by the opposite-spin and -momentum pairs which form at zero temperature can be produced by means of laser techniques similar to the one sketched in the previous paragraph \cite{Miller2007}. For what concerns $g(n)$, an expression for which as a functional of the longitudinal pair density $n(x,t)$ was notably obtained in the unitary limit using some Gaussian variational ansatz for the radial wavefunction of the atomic cloud \cite{Adhikari2009}. This expression simplifies to $g(n)=C_{1}\hbar\omega_{\perp}(na_{\perp})^{2/3}$ when $na_{\perp}\ll1$ and to $g(n)=C_{2}\hbar\omega_{\perp}(na_{\perp})^{2/5}$ when $na_{\perp}\gg1$, where $C_{1}\simeq2.968$ and $C_{2}\simeq2.940$ are constants of which only the numerical values are given here and $a_{\perp}=[2\hbar/(m\omega_{\perp})]^{1/2}$ is the oscillator length of the transverse harmonic confinement felt by the atoms. The flow of the Fermi fluid in the positive-$x$ direction can be simulated in the same manner as the one described in the previous paragraph \cite{Miller2007}.

Equation~\eqref{Eq:GNLSE} is also encountered in the optical realm. For instance, it phenomenologically models the evolution of a condensate of exciton-polaritons of effective mass $m$ in a semiconductor optical microcavity designed in the form of a 1D guide of $x$ axis, disregarding possible polarization effects of the light modes in the cavity \cite{Carusotto2013}. In this context, the obstacle potential $U(x)$ typically results from the presence of a structural defect in the device \cite{Amo2009}. It can also be artificially created by means of lithographic techniques or using a CW laser. On the other hand, there is a whole body of evidence showing that the overall effective interaction potential $g(n)$ between exciton-polaritons is repulsive and linearly scales with the exciton-polariton density $n(x,t)$ \cite{Estrecho2019}. Importantly, the last term in the right-hand side of Eq.~\eqref{Eq:GNLSE} describes the effect of cavity losses due to the finiteness of the exciton-polariton lifetime, proportional to $1/\gamma$. To maintain the system in time, the latter then needs to be optically pumped. Not included in Eq.~\eqref{Eq:GNLSE}, both phenomenological and microscopic pumping schemes are reviewed in Ref.~\cite{Carusotto2013}. In many experiments, the exciton-polariton condensate is put into motion with respect to the obstacle using a resonant pump in nonzero incidence with the sample \cite{Amo2009}.

Always in optics and as a last example, Eq.~\eqref{Eq:GNLSE} is also used to describe the propagation of a scalar laser field in a local nonlinear medium. Its applications range from both fundamental and applied nonlinear dynamics \cite{Boyd2020, Agrawal2019} to the simulation of atomic physics with light \cite{Leboeuf2010, Larre2015b, Vocke2016, Santic2018, Michel2018, Cherroret2018, Glorieux2018, Eloy2021}. In this classical-optics framework, the reduced Planck constant is absent from Eq.~\eqref{Eq:GNLSE} (formally, $\hbar=1$), $t=z$ is the propagation coordinate of the laser carrier, $m$ is its propagation constant (or the opposite inverse of the group-velocity dispersion in fiber optics), $x$ is one of the two paraxial coordinates (or time), and $U(x)$ describes some spatial (or temporal) inhomogeneity in the linear refractive index invariant along the $z$ axis \cite{Michel2018, Eloy2021}. In the latter experiments, the 1D regime of Eq.~\eqref{Eq:GNLSE} could be obtained by shaping both the laser and the index defect in a way that they are invariant along the $y$ axis. Regarding $g(n)$, it linearly increases with the light intensity $n(x,t)$ in a defocusing Kerr medium \cite{Boyd2020, Agrawal2019}. In a defocusing saturable medium like the nonlinear photorefractive crystal used in the experiments of Refs.~\cite{Michel2018, Eloy2021}, it is of the form $g(n)=\pi N^{3}r_{33}E_{0}n/[\lambda_{0}(n_{\mathrm{s}}+n)]$, where $N$ and $r_{33}$ are respectively the mean refractive index and the electro-optic coefficient of the crystal along the extraordinary axis, $E_{0}$ is the amplitude of an electric field applied to the crystal along the $\mathrm{c}$-axis, $\lambda_{0}$ is the wavelength of the laser carrier in free space, and $n_{\mathrm{s}}$ is a saturation intensity adjusted by illuminating the crystal with white light. It is worth noting that light suffers from linear (mainly) absorption in such a material. This is described by the $\gamma$ term in Eq.~\eqref{Eq:GNLSE}, proportional to the imaginary part of the linear electric susceptibility \cite{Boyd2020, Agrawal2019}. Finally, by tilting the laser at a small angle to the index defect, we optically simulate the flow of an atomic quantum fluid past an obstacle \cite{Michel2018, Eloy2021}.

\section{Derivation of Eqs.~\texorpdfstring{\eqref{Eq:VcNarrow}}{Lg} and \texorpdfstring{\eqref{Eq:N0cNarrow}}{Lg}}
\label{App:B}

In this appendix, we detail the calculations leading to Eqs.~\eqref{Eq:VcNarrow} and \eqref{Eq:N0cNarrow} giving the critical velocity for superfluidity $v_{\mathrm{c}}$ in the case of an obstacle potential of the form $U(x)=U_{0}F(\sigma)\delta(x)$ [$U(x)=U_{0}f(|x|/\sigma)$ in the limit $\sigma\ll1$, with $F(\sigma)$ being the integral of $f(|x|/\sigma)$ over the whole real axis]. These calculations are inspired by previous studies \cite{Hakim1997, Leboeuf2001, Pavloff2002} which mostly focus on the standard nonlinear Schr\"odinger potential $g(n)=n$.

\begin{figure}[t!]
\includegraphics[width=\linewidth]{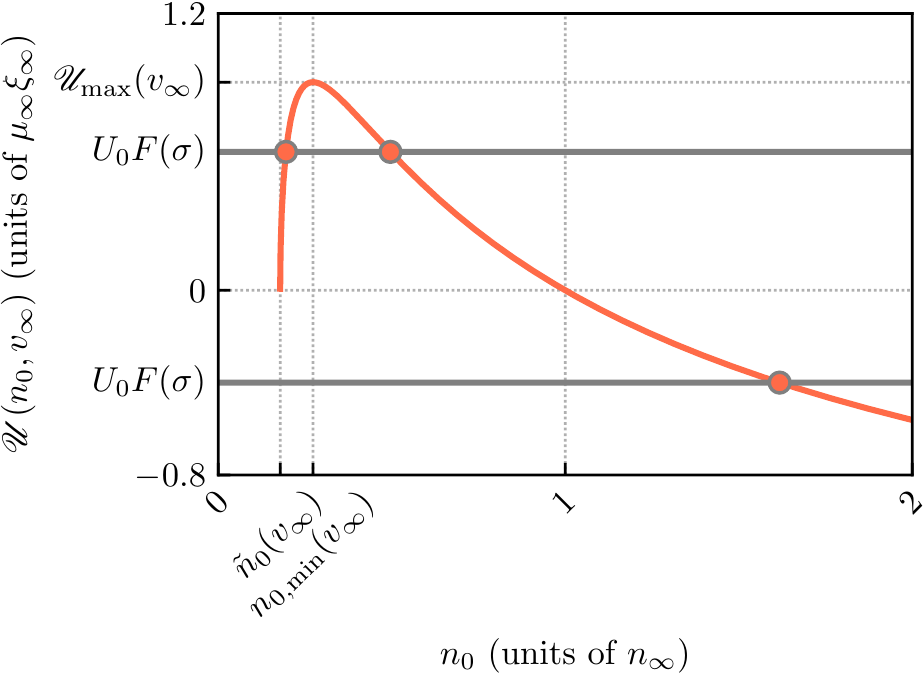}
\caption{The curve represents $\mathscr{U}(n_{0},v_{\infty})$ as a function of $n_{0}$ [Eq.~\eqref{Eq:UNarrow}] for $v_{\infty}=0.5$ and $g(n)=(1+n_{\mathrm{s}})^{2}n/[n_{\mathrm{s}}(n_{\mathrm{s}}+n)]$ with $n_{\mathrm{s}}=1$. In this case, $\tilde{n}_{0}(v_{\infty})\simeq0.178$, $n_{0,\mathrm{min}}(v_{\infty})\simeq0.273$, and $\mathscr{U}_{\mathrm{max}}(v_{\infty})\simeq0.902$. When $U_{0}F(\sigma)>0$ [upper line, $U_{0}F(\sigma)=0.6$ on the plot], the graph shows that $\mathscr{U}(n_{0},v_{\infty})=U_{0}F(\sigma)$ admits solutions (abscissas of the two upper markers) provided $U_{0}F(\sigma)<\mathscr{U}_{\mathrm{max}}(v_{\infty})$. The physical solution is the largest because it is the only one approaching $1$ as $U_{0}F(\sigma)\to0$, in accordance with perturbation theory. When $U_{0}F(\sigma)<0$ [lower line, $U_{0}F(\sigma)=-0.4$ on the plot], the graph suggests that $\mathscr{U}(n_{0},v_{\infty})=U_{0}F(\sigma)$ always admits a solution (abscissa of the lower marker).}
\label{Fig:GeneralizedPotential_Narrow}
\end{figure}

For such a $\delta$-peaked obstacle, the solutions of Eq.~\eqref{Eq:HydroEqDensityBis} going to unity at both infinities are given by $n'(x)=\mathrm{sgn}(x)2n\mathscr{U}(n,v_{\infty})$ for all $x\neq0$ [Eq.~\eqref{Eq:HydroEqDensityBis} with $U(x)=0$ integrated over the whole real axis], where
\begin{align}
\notag
\mathscr{U}(n,v_{\infty})&\left.=\mathrm{sgn}(1-n)\sqrt{2}\bigg[{-}\frac{v_{\infty}^{2}}{2}\bigg(1-\frac{1}{n}\bigg)^{2}\right. \\
\label{Eq:UNarrow}
&\left.\hphantom{=}-g(1)+\frac{g(1)-G(1)+G(n)}{n}\bigg]^{\frac{1}{2}},\right.
\end{align}
with $G(n)=\int dn\,g(n)$. We match these solutions at $x=0$ with $[n'(0^{+})-n'(0^{-})]/(4n_{0})=U_{0}F(\sigma)$ [Eq.~\eqref{Eq:HydroEqDensityBis} integrated over an infinitesimal-length interval containing the origin], where $n_{0}=n(0)$. Everything combined yields $\mathscr{U}(n_{0},v_{\infty})=U_{0}F(\sigma)$ \cite{NoteSign}, the condition of existence of the solutions $n_{0}$ of which defines $v_{\mathrm{c}}$. To understand it, we plot $\mathscr{U}(n_{0},v_{\infty})$ as a function of $n_{0}$ at a given $v_{\infty}$ and focus on its intersections with $U_{0}F(\sigma)\gtrless0$. This is done in Fig.~\ref{Fig:GeneralizedPotential_Narrow}
for $v_{\infty}=0.5$ and $g(n)=(1+n_{\mathrm{s}})^{2}n/[n_{\mathrm{s}}(n_{\mathrm{s}}+n)]$ with $n_{\mathrm{s}}=1$, but nothing qualitatively changes for other $v_{\infty}$'s and $g(n)$'s.

When $U_{0}>0$ (repulsive obstacle potential), $U_{0}F(\sigma)$ always intersects $\mathscr{U}(n_{0},v_{\infty})$ provided it is smaller than the maximum $\mathscr{U}_{\mathrm{max}}(v_{\infty})$ of $\mathscr{U}(n_{0},v_{\infty})$, reached at a $n_{0}=n_{0,\mathrm{min}}(v_{\infty})$ solution of $\mathscr{U}_{n_{0}}[n_{0,\mathrm{min}}(v_{\infty}),v_{\infty}]=0$, i.e.,
\begin{align}
\notag
\frac{n_{0,\mathrm{min}}(v_{\infty})}{1-n_{0,\mathrm{min}}(v_{\infty})}\{g(1)&-G(1)-g[n_{0,\mathrm{min}}(v_{\infty})]n_{0,\mathrm{min}}(v_{\infty}) \\
\label{Eq:N0cNarrowApp}
&+G[n_{0,\mathrm{min}}(v_{\infty}^{\vphantom{2}})]\}=v_{\infty}^{2}.
\end{align}
Since $\mathscr{U}_{\mathrm{max}}(v_{\infty})$ is a decreasing function of $v_{\infty}$ (see below), $U_{0}F(\sigma)<\mathscr{U}_{\mathrm{max}}(v_{\infty})$ equivalently reads $v_{\infty}<v_{\mathrm{c}}$, where $v_{\mathrm{c}}$ is a function of $U_{0}F(\sigma)$ implicitly given by
\begin{equation}
\label{Eq:VcNarrowApp}
\mathscr{U}_{\mathrm{max}}(v_{\mathrm{c}})=U_{0}F(\sigma).
\end{equation}
Equations~\eqref{Eq:UNarrow}--\eqref{Eq:VcNarrowApp} are basically Eqs.~\eqref{Eq:VcNarrow} and \eqref{Eq:N0cNarrow}. No simple expression for $\mathscr{U}_{\mathrm{max}}(v_{\mathrm{c}})$ exists in the case of the $g(n)$'s considered in this work, except when $g(n)=n$ for which Eq.~\eqref{Eq:VcNarrowApp} reads \cite{Hakim1997, Leboeuf2001, Pavloff2002}
\begin{equation}
\label{Eq:VcNarrowGPE}
\frac{[1-20v_{\mathrm{c}}^{2}-8v_{\mathrm{c}}^{4}+(1+8v_{\mathrm{c}}^{2})^{\frac{3}{2}}]^{\frac{1}{2}}}{2\sqrt{2}v_{\mathrm{c}}}=U_{0}F(\sigma).
\end{equation}
The fact that $\mathscr{U}_{\mathrm{max}}(v_{\infty})$ decreases with $v_{\infty}$ can be checked in a graphical way and understood as follows: When the fluid velocity increases, kinetic effects gradually prevail over interaction-induced phenomena like superfluidity, the breakdown of which is prevented by an effective reduction of the perturbation potential. At last, it is worth noting that there are two possible solutions $n_{0}$ to $\mathscr{U}(n_{0},v_{\infty})=U_{0}F(\sigma)$ (see Fig.~\ref{Fig:GeneralizedPotential_Narrow}). In fact, the largest is the physical one. Indeed, the latter is the only one approaching $1$ as $U_{0}F(\sigma)\to0$, in accordance with perturbation theory. This indicates that at a fixed $v_{\infty}$, $n_{0,\mathrm{min}}(v_{\infty})$ corresponds to the smallest possible value of the density at $x=0$, hence the subscript ``$\mathrm{min}$.'' It is reached when $v_{\infty}=v_{\mathrm{c}}$, denoted by $n_{0,\mathrm{c}}$ in the main text.

Finally, when $U_{0}<0$ (attractive obstacle potential), Fig.~\ref{Fig:GeneralizedPotential_Narrow} suggests that there is no lower bound on $U_{0}F(\sigma)$ for it to intersect $\mathscr{U}(n_{0},v_{\infty})$. This means that the flow can be superfluid for all $v_{\infty}<1$, hence $v_{\mathrm{c}}=1$ for all $U_{0}F(\sigma)$. This result is rigorously valid for continuously increasing $g(n)$'s, e.g., for $g(n)=n^{\nu}/\nu$ with $\nu>0$. In the case of the saturable $g(n)$ used in Fig.~\ref{Fig:GeneralizedPotential_Narrow}, $\mathscr{U}(n,v_{\infty})$ given in Eq.~\eqref{Eq:UNarrow} tends to $-(-v_{\infty}^{2}+2+2n_{\mathrm{s}})^{1/2}$ at large density (at constant $v_{\infty}$ and $n_{\mathrm{s}}$). Superfluidity for all $v_{\infty}<1$ is thus possible provided $U_{0}F(\sigma)>-(1+2n_{\mathrm{s}})^{1/2}$. This turns out to be verified for a very wide range of negative values of $U_{0}F(\sigma)$ provided $n_{\mathrm{s}}$ is large, which is actually the case in the experiment of Ref.~\cite{Eloy2021} (see footnote \cite{NoteNonlinearity}).

\section{Derivation of Eqs.~\texorpdfstring{\eqref{Eq:VcWide}}{Lg}--\texorpdfstring{\eqref{Eq:N0cWide0}}{Lg}}
\label{App:C}

Up to second order in $1/\sigma\ll1$, one has $U(x)=U_{0}[1+f''(0)x^{2}/(2\sigma^{2})]$. Given such a quasi-flat potential, it is natural to assume that $n(x)$ weakly deviates from $n_{0}$ with derivatives scaling as positive powers of $1/\sigma$. In this appendix, we detail the multiple-scale treatment of Eq.~\eqref{Eq:HydroEqDensityBis} leading to Eqs.~\eqref{Eq:VcWide}--\eqref{Eq:N0cWide0}. To begin with, we derive the critical velocity for superfluidity $v_{\mathrm{c}}$ as resulting from the first, $\sigma$-independent term in the expansion of $U(x)$ written above. Then, we account for the second, $1/\sigma^{2}$ term and perturbatively obtain the leading $\sigma$ dependence of $v_{\mathrm{c}}$. Our calculations generalize the ones performed in Ref.~\cite{Hakim1997} for $g(n)=n$.

\begin{figure}[t!]
\includegraphics[width=\linewidth]{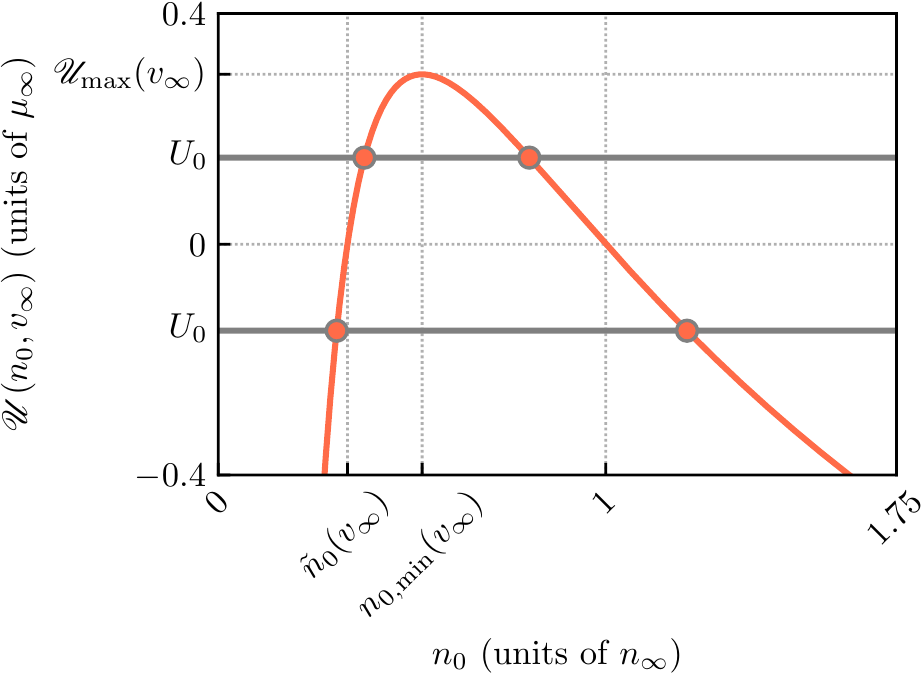}
\caption{Same as Fig.~\ref{Fig:GeneralizedPotential_Narrow} but for $\mathscr{U}(n_{0},v_{\infty})$ given by Eq.~\eqref{Eq:UWide}. In this case, $\tilde{n}_{0}(v_{\infty})=1/3$, $n_{0,\mathrm{min}}(v_{\infty})\simeq0.526$, and $\mathscr{U}_{\mathrm{max}}(v_{\infty})\simeq0.294$. The values of $U_{0}$ considered on the plot are $U_{0}=0.15$ (upper line) and $U_{0}=-0.15$ (lower line).}
\label{Fig:GeneralizedPotential_Wide}
\end{figure}

At zeroth order in $1/\sigma$, $U(x)=U_{0}$ and $n(x)=n_{0}$ for all $x$. In this case, defining
\begin{equation}
\label{Eq:UWide}
\mathscr{U}(n,v_{\infty})=\frac{v_{\infty}^{2}}{2}\bigg(1-\frac{1}{n^{2}}\bigg)+g(1)-g(n),
\end{equation}
Eq.~\eqref{Eq:HydroEqDensityBis} reduces to the simple algebraic equation $\mathscr{U}(n_{0},v_{\infty})=U_{0}$. This equation corresponds to the zeroth order of the so-called hydraulic approximation of Eq.~\eqref{Eq:HydroEqDensityBis} (see, e.g., Ref.~\cite{Leszczyszyn2009}). Under what condition on $v_{\infty}$ do the solutions $n_{0}$ of this equation, hence superfluidity, exist? To answer this question, we reason in the same way as in Appendix~\ref{App:B} from the plot of $\mathscr{U}(n_{0},v_{\infty})$ as a function of $n_{0}$ at a fixed $v_{\infty}$. Figure~\ref{Fig:GeneralizedPotential_Wide}
shows it for the same value of $v_{\infty}$ and the same $g(n)$ as the ones used in Fig.~\ref{Fig:GeneralizedPotential_Narrow}.

When $U_{0}>0$ (repulsive obstacle potential), the critical velocity $\bar{v}_{\mathrm{c}}$ is given by an equation similar to Eq.~\eqref{Eq:VcNarrowApp}:
\begin{equation}
\label{Eq:VcWide0App}
\mathscr{U}_{\mathrm{max}}(\bar{v}_{\mathrm{c}})=U_{0},
\end{equation}
where the maximum $\mathscr{U}_{\mathrm{max}}(v_{\infty})$ of $\mathscr{U}(n_{0},v_{\infty})$ is reached at a $n_{0}=n_{0,\mathrm{min}}(v_{\infty})$ solution of $\mathscr{U}_{n_{0}}[n_{0,\mathrm{min}}(v_{\infty}),v_{\infty}]=0$, i.e.,
\begin{equation}
\label{Eq:N0cWide0App}
g'[n_{0,\mathrm{min}}^{\vphantom{3}}(v_{\infty}^{\vphantom{2}})]n_{0,\mathrm{min}}^{3}(v_{\infty}^{\vphantom{2}})=v_{\infty}^{2}.
\end{equation}
Equations~\eqref{Eq:UWide}--\eqref{Eq:N0cWide0App} are basically Eqs.~\eqref{Eq:VcWide0} and \eqref{Eq:N0cWide0}. Equation~\eqref{Eq:VcWide0App} is especially simple in the case where $g(n)=n^{\nu}/\nu$ \cite{Albert2010}:
\begin{equation}
\label{Eq:VcWide0PowerLaw}
\frac{1}{\nu}-\frac{1}{2}\bigg(\frac{2+\nu}{\nu}\bar{v}_{\mathrm{c}}^{\frac{2\nu}{2+\nu}}-\bar{v}_{\mathrm{c}}^{2\vphantom{\frac{2\nu}{2+\nu}}}\bigg)=U_{0},
\end{equation}
the $\nu=1$ version of which may be found in, e.g., Refs.~\cite{Hakim1997, Leszczyszyn2009}.

When $U_{0}<0$ (repulsive obstacle potential), we find again that $\bar{v}_{\mathrm{c}}=1$ for all $U_{0}$ (more precisely for all $U_{0}>-(1+2n_{\mathrm{s}})/2$ in the case of the saturable $g(n)$ used in Fig.~\ref{Fig:GeneralizedPotential_Wide}, the right-hand side of this inequality being still very negative in a recent experiment \cite{Eloy2021, NoteNonlinearity}). Compared to Fig.~\ref{Fig:GeneralizedPotential_Narrow}, two $n_{0}$'s are possible when $U_{0}<0$. The physical one is the largest for the same reason as the one invoked in Appendix~\ref{App:B}.

We now account for the $1/\sigma^{2}$ term in the series expansion of $U(x)$ written above and search for the corresponding critical velocity $v_{\mathrm{c}}=\bar{v}_{\mathrm{c}}+\delta v_{\mathrm{c}}$, where $\delta v_{\mathrm{c}}$ is a small correction to $\bar{v}_{\mathrm{c}}$ which must depend on $U_{0}$, $f''(0)$, and $\sigma$. We restrict the calculations to the case of a repulsive obstacle potential [$U_{0}>0$ and $f''(0)<0$], for which $\bar{v}_{\mathrm{c}}$ is not trivial. We start by considering small departures of $v_{\infty}$ and $n(x)$ from $\bar{v}_{\mathrm{c}}$ and $n_{0}=n_{0,\mathrm{min}}(\bar{v}_{\mathrm{c}})=\bar{n}_{0,\mathrm{c}}$, respectively: $v_{\infty}=\bar{v}_{\mathrm{c}}+\delta v_{\infty}$ and $n(x)=\bar{n}_{0,\mathrm{c}}+\delta n(x)$. After substitution into Eq.~\eqref{Eq:HydroEqDensityBis}, we get, at the first nonzero order in $\delta n(x)$ and $\delta v_{\infty}$, the following nonlinear differential equation for $\delta n(x)$:
\begin{equation}
\label{Eq:HydroEqDensityTer}
\delta n''-2\alpha\delta n^{2}=-\frac{2\beta}{\sigma^{2}}x^{2}+4\gamma\delta v_{\infty},
\end{equation}
where $\alpha=3\bar{v}_{\mathrm{c}}^{2}/\bar{n}_{0,\mathrm{c}}^{3}+g''(\bar{n}_{0,\mathrm{c}}^{\vphantom{3}})\bar{n}_{0,\mathrm{c}}^{\vphantom{3}}>0$, $\beta=U_{0}|f''(0)|\bar{n}_{0,\mathrm{c}}$, and $\gamma=\bar{v}_{\mathrm{c}}(1/\bar{n}_{0,\mathrm{c}}-\bar{n}_{0,\mathrm{c}})$.

A trivial scale analysis of Eq.~\eqref{Eq:HydroEqDensityTer} shows that $x=O(\sigma^{1/3})$, $\delta n(x)=O(1/\sigma^{2/3})$, and $\delta v_{\infty}=O(1/\sigma^{4/3})$. This makes it possible to rescale it in the form
\begin{equation}
\label{Eq:HydroEqDensityQuater}
\Delta n''-\Delta n^{2}=-X^{2}+\Delta v_{\infty},
\end{equation}
where we have defined $X=2^{1/3}\alpha^{1/6}\beta^{1/6}\times x/\sigma^{1/3}$, $\Delta n(X)=2^{1/3}\alpha^{2/3}\beta^{-1/3}\times\sigma^{2/3}\delta n(x)$, and $\Delta v_{\infty}=2^{5/3}\alpha^{1/3}\beta^{-2/3}\gamma\times\sigma^{4/3}\delta v_{\infty}$. Implementing a relaxation algorithm similar to the one sketched in Sec.~\ref{Sec:ObstacleOfArbitraryWidth}, we numerically find that the solutions $\Delta n(X)$ of Eq.~\eqref{Eq:HydroEqDensityQuater} exist as long as $-\infty<\Delta v_{\infty}<C\simeq1.466$, which was also established in Ref.~\cite{Hakim1997}. We then infer that the first nonzero correction to $\bar{v}_{\mathrm{c}}$ reads
\begin{equation}
\label{Eq:VcWideApp}
\delta v_{\mathrm{c}}=\frac{C}{2^{\frac{5}{3}}}\frac{\beta^{\frac{2}{3}}}{\alpha^{\frac{1}{3}}\gamma}\frac{1}{\sigma^{\frac{4}{3}}},
\end{equation}
which is nothing but $v_{\mathrm{c}}-\bar{v}_{\mathrm{c}}$ given in Eq.~\eqref{Eq:VcWide}.

\section{Derivation of Eq.~\texorpdfstring{\eqref{Eq:VcJosephson}}{Lg}}
\label{App:D}

In this appendix, we evaluate the (conserved) current density $j=\mathrm{Im}[\psi^{\ast}(x)\psi'(x)]$ associated with the solution $\psi(x)$ of the nondissipative and stationary version of Eq.~\eqref{Eq:GNLSE},
\begin{equation}
\label{Eq:StationaryGNLSE}
\mu\psi=-\frac{1}{2}\psi''+U(x)\psi+g(|\psi|^{2})\psi,
\end{equation}
in the case where the obstacle potential $U(x)=U_{0}f(|x|/\sigma)$ is both strongly repulsive (large $U_{0}>0$) and slowly varying (large $\sigma$). Units are those used in the main text.

In the presence of such a large-amplitude potential, $j$ must be very small. Therefore, it is possible to obtain its leading behavior without accounting for the flux at infinity in its derivation. In this approximation, one has $\mu=g(1)$ for the energy of the stationary state of the fluid. Regarding the corresponding wavefunction $\psi(x)$, we search for it in the form
\begin{equation}
\label{Eq:Ansatz}
\psi(x)=\psi_{\mathrm{L}}(x)\exp(i\theta_{\mathrm{L}})+\psi_{\mathrm{R}}(x)\exp(i\theta_{\mathrm{R}}),
\end{equation}
where the ``$\mathrm{L}$'' (``$\mathrm{R}$'') component, solution of Eq.~\eqref{Eq:StationaryGNLSE} with a $x$-dependent modulus $\psi_{\mathrm{L}}(x)$ [$\psi_{\mathrm{R}}(x)$] and a constant phase $\theta_{\mathrm{L}}$ ($\theta_{\mathrm{R}}$), describes the fluid at left (right) of its density dip, that is, for $x<0$ ($x>0$). In the following, we mainly focus on the left solution (the right solution is straightforwardly deduced by symmetry) and denote by $-x_{\mathrm{cl}}<0$ the classical turning point of the fluid in this region of space: $U(-x_{\mathrm{cl}})=\mu$ ($x_{\mathrm{cl}}$ is its classical turning point for $x>0$). The calculations below are inspired by Ref.~\cite{Dalfovo1996}.

Deep in the classical, $x\ll-x_{\mathrm{cl}}$ region where $\mu>U(x)$, $U(x)$ varies so slowly that the kinetic term in Eq.~\eqref{Eq:StationaryGNLSE} can be neglected (Thomas-Fermi approximation). In this case, $\psi_{\mathrm{L}}(x)$ is implicitly given by
\begin{equation}
\label{Eq:ThomasFermi}
g[\psi_{\mathrm{L}}^{2}(x)]=\mu-U(x).
\end{equation}
Deep in the classically forbidden, $x\gg-x_{\mathrm{cl}}$ region where $\mu<U(x)$, $U(x)$ is so strong that this time it is the nonlinear term in Eq.~\eqref{Eq:StationaryGNLSE} that can be neglected. In this case, $\psi_{\mathrm{L}}(x)$ obeys the Schr\"odinger equation $\mu\psi_{\mathrm{L}}^{\vphantom{\prime\prime}}=-\psi_{\mathrm{L}}''/2+U(x)\psi_{\mathrm{L}}^{\vphantom{\prime\prime}}$. In the limit $\sigma\gg1$, this equation can be solved in the coordinate $\tilde{x}=x/\sigma$ by the ansatz $\psi_{\mathrm{L}}(\tilde{x})=\exp[\sigma\int d\tilde{x}\,k(\tilde{x})]$ with $k(\tilde{x})=\sum_{\ell\geqslant0}k_{\ell}(\tilde{x})/\sigma^{\ell}$. To $\ell=1$ order, this gives the well-known WKB (or semiclassical) solution
\begin{equation}
\label{Eq:WKB}
\psi_{\mathrm{L}}(x)=\frac{\alpha}{2^{\frac{1}{4}}}\frac{\exp({-}\sqrt{2}\int_{-x_{\mathrm{cl}}}^{x}dx'\,[U(x')-\mu]^{\frac{1}{2}})}{[U(x)-\mu]^{\frac{1}{4}}},
\end{equation}
where $\alpha$ is an integration constant which depends on the parameters of the obstacle potential. We determine it at the end of this appendix in the case where $g(|\psi|^{2})=|\psi|^{2\nu}/\nu$, for which the result is explicit. By symmetry, one has $\psi_{\mathrm{R}}(x)=\psi_{\mathrm{L}}(-x)$ for all $x>0$.

The Thomas-Fermi solution \eqref{Eq:ThomasFermi} and its symmetric for $x>0$ contribute negligibly to the current density since they are slowly varying functions of $x$. Hence, $j$ is mainly ruled by the WKB solutions $\psi_{\mathrm{L}}(x)$ and $\psi_{\mathrm{R}}(x)=\psi_{\mathrm{L}}(-x)$ given by Eq.~\eqref{Eq:WKB}, i.e., by the wavefunction of the fluid in the classically forbidden, $-x_{\mathrm{cl}}<x<x_{\mathrm{cl}}$ region. Straightforward manipulations yield the Josephson-type relation
\begin{equation}
\label{Eq:Josephson}
j=j_{\mathrm{c}}\sin(\theta_{\mathrm{R}}-\theta_{\mathrm{L}}),
\end{equation}
where $j_{\mathrm{c}}=\psi_{\mathrm{L}}^{\vphantom{\prime}}(x)\psi_{\mathrm{R}}'(x)-\psi_{\mathrm{L}}'(x)\psi_{\mathrm{R}}^{\vphantom{\prime}}(x)$ is indeed $x$ independent, given by
\begin{equation}
\label{Eq:VcJosephsonApp}
j_{\mathrm{c}}=2\alpha^{2}\exp\!\bigg\{{-}\sqrt{2}\int_{-x_{\mathrm{cl}}}^{x_{\mathrm{cl}}}dx\,[U(x)-\mu]^{\frac{1}{2}}\bigg\}.
\end{equation}
Since the dimensionless quantities $j$ and $v_{\infty}$ coincide [the current density is also expressed as $j=n(x)v(x)$ and is conserved], the maximum current density \eqref{Eq:VcJosephsonApp} for tunneling across the barrier is nothing but the critical velocity $v_{\mathrm{c}}$ for superfluidity past the obstacle. By defining $A=2\alpha^{2}$, we end up with Eq.~\eqref{Eq:VcJosephson} of the main text.

In this concluding paragraph, we establish the dependence of the integration constant $\alpha$ on the parameters of the obstacle potential by numerically matching Eq.~\eqref{Eq:ThomasFermi} to Eq.~\eqref{Eq:WKB} in a neighbourhood of $-x_{\mathrm{cl}}$ where $U(x)=\mu+U'(-x_{\mathrm{cl}})(x+x_{\mathrm{cl}})+o(x+x_{\mathrm{cl}})$ is in good approximation an increasing linear function of $x$. We specifically focus on the nonlinear potential $g(|\psi|^{2})=|\psi|^{2\nu}/\nu$ for which the result is explicit. In this neighbourhood of $-x_{\mathrm{cl}}$ and for this nonlinearity, Eq.~\eqref{Eq:StationaryGNLSE} for the left wavefunction can be expressed in the form
\begin{equation}
\label{Eq:BoundaryLayer}
\phi_{\mathrm{L}}^{\prime\prime\vphantom{1+2\nu}}-\chi\phi_{\mathrm{L}}^{\vphantom{1+2\nu}}-\phi_{\mathrm{L}}^{1+2\nu}=0,
\end{equation}
where $\chi=(x+x_{\mathrm{cl}})/\delta$ and $\phi_{\mathrm{L}}(\chi)=(2\delta^{2}/\nu)^{1/(2\nu)}\psi_{\mathrm{L}}(x)$ with $\delta=[2U'(-x_{\mathrm{cl}})]^{-1/3}$. According to what precedes, when $\chi\to-\infty$, $\phi_{\mathrm{L}}(\chi)$ should be given by the Thomas-Fermi solution of Eq.~\eqref{Eq:BoundaryLayer}:
\begin{equation}
\label{Eq:ThomasFermiBoundaryLayer}
\phi_{\mathrm{L}}(\chi)=(-\chi)^{1/(2\nu)}.
\end{equation}
On the other hand, when $\chi\to+\infty$, $\phi_{\mathrm{L}}(\chi)$ should be solution of the linear version of Eq.~\eqref{Eq:BoundaryLayer}, which is the well-known Airy differential equation whose solution $\alpha_{\nu}\mathrm{Ai}(\chi)$ asymptotically behaves as
\begin{equation}
\label{Eq:WKBBoundaryLayer}
\phi_{\mathrm{L}}(\chi)=\frac{\alpha_{\nu}}{2\sqrt{\pi}}\frac{\exp(-\frac{2}{3}\chi^{\frac{3}{2}})}{\chi^{\frac{1}{4}}},
\end{equation}
where $\alpha_{\nu}$ is a constant which we adjust so that the numerical solution of Eq.~\eqref{Eq:BoundaryLayer} interpolates the asymptotic behaviors \eqref{Eq:ThomasFermiBoundaryLayer} and \eqref{Eq:WKBBoundaryLayer}. For example, we find $\alpha_{1}\simeq1.407$ and $\alpha_{1/2}\simeq1.306$ for the two nonlinearities of the second paragraph of Appendix~\ref{App:A}. By making Eqs.~\eqref{Eq:WKB} and \eqref{Eq:WKBBoundaryLayer} coincide at the right border of the neighbourhood of $-x_{\mathrm{cl}}$ [this is easily done by recognizing that the argument of the exponential in Eq.~\eqref{Eq:WKBBoundaryLayer} is the opposite of the antiderivative of $\sqrt{\chi}$], we eventually get
\begin{equation}
\label{Eq:Alpha}
\alpha=\frac{\nu^{\frac{1}{2\nu}}\alpha_{\nu}}{2^{\frac{1+5\nu}{6\nu}}\sqrt{\pi}}U'(-x_{\mathrm{cl}})^{\frac{2+\nu}{6\nu}},
\end{equation}
hence $A\propto U'(-x_{\mathrm{cl}})^{(2+\nu)/(3\nu)}$ for the amplitude of the exponential in Eq.~\eqref{Eq:VcJosephsonApp}.


\begin{thebibliography}{99}
\bibitem{NoteAsymptotics}
To have superfluidity, it is not essential that the density (velocity) of the fluid tends to the same constant value $n_{\infty}$ ($v_{\infty}$) at both infinities. Nevertheless but without too much loss of generality, we restrict ourselves to this case for the sake of simplicity.
\bibitem{NoteSign}
Given this constraint on $n_{0}$, $\mathrm{sgn}(1-n)$ in Eq.~\eqref{Eq:UNarrow} indicates that $n(x)$ displays a hole or a bump around $x=0$ depending on whether $U_{0}>0$ or $U_{0}<0$, which is intuitively expected. Concretely, when $g(n)=n$ for example, one has $n(x)=v_{\infty}^{2}+(1-v_{\infty}^{2})\tanh^{\mathrm{sgn}(U_{0})2}[(1-v_{\infty}^{2})^{1/2}(|x|+x_{0})]$, where $x_{0}[v_{\infty},U_{0}F(\sigma)]>0$ is determined from the discontinuity of $n'(x)$ at $x=0$. For other $g(n)$'s, $n(x)$ obviously differs from this solitonic solution---and may not be explicit---but qualitatively behaves in the same way.
\end{thebibliography}
\end{document}